\documentclass[preprint]{aastex}
\usepackage{natbib}
\usepackage{rotating}
\usepackage{lscape}
\newcommand{\allframe}{{\sc allframe}}
\newcommand{\daophot}{{\sc daophot}}
\newcommand{\allstar}{{\sc allstar}}
\newcommand{\mopex}{{\sc mopex}}

\slugcomment{Draft last edited \today}

\shorttitle{IC~1613 Cepheids}
\shortauthors{Scowcroft et al.}

\begin{document}


%\title{The Carnegie Hubble Program: \\
 %   The Leavitt law at 3.6 $\mu$ m and 4.5 $\mu$ m in IC~1613}
 
 \title{The Carnegie Hubble Program: The Infrared Leavitt Law in IC~1613}


\author{\bf Victoria Scowcroft, Wendy L. Freedman, Barry F. Madore, Andrew J. Monson, S.~E.~Persson, Mark Seibert,}
\affil{Observatories of the Carnegie Institution for Science \\ 813
Santa Barbara St., Pasadena, CA~91101}
\author{\bf Jane R. Rigby}
\affil{Observational Cosmology Lab, NASA Goddard Space Flight Center, Greenbelt MD 20771}
\author{\bf \&}
\author{\bf Jason Melbourne}
\affil{Caltech Optical Observatories, Division of Physics, Mathematics and Astronomy, Mail Stop 301-17, California Institute of Technology, Pasadena, CA 91125}
\email{vs@obs.carnegiescience.edu, wendy@obs.carnegiescience.edu, 
barry@obs.carnegiescience.edu, amonson@obs.carnegiescience.edu,
persson@obs.carnegiescience.edu,
mseibert@obs.carnegiescience.edu,
jane.r.rigby@nasa.gov
jmel@caltech.edu} 


%\altaffiltext{1}{Alternative}
%\altaffiltext{2}{affiliations}
%\altaffiltext{4}{go}
%\altaffiltext{5}{here}


\begin{abstract}
We have observed the dwarf galaxy IC~1613, at multiple epochs in the mid--infrared using \textit{Spitzer} and contemporaneously in the near--infrared using the new FourStar near-IR camera on Magellan. We have constructed Cepheid period--luminosity relations in the $J$, $H$, $K_{s}$, $[3.6]$ and $[4.5]$ bands and have used  %their observed zero--points 
the run of their apparent distance moduli as a function of wavelength to derive the line--of--sight reddening and distance to IC~1613. Using a nine--band fit, we find $E(B-V) = 0.05 \pm 0.01$~mag and an extinction--corrected distance modulus of $\mu_{0} = 24.29 \pm 0.03_{statistical} \pm 0.03_{systematic}$~mag. By comparing our multi--band and $[3.6]$ distance moduli to results from the tip of the red giant branch and red clump distance indicators, we find that metallicity has no measurable effect on Cepheid distances at 3.6 $\mu$m in the metallicity range $-1.0 \leq [Fe/H] \leq 0.2$, hence derivations of the Hubble constant at this wavelength require no correction for metallicity. %This result is in agreement with that of Udalski et al. (2001).

%no correction or systematic uncertainty for metallicity. 
\end{abstract}



\section{Introduction}
\label{sec:introduction}
 
The stated aim of the Carnegie Hubble Program (CHP) is to measure the Hubble constant to an accuracy of 2\% using data from the Warm \textit{Spitzer} mission, and future data from \textit{JWST} and Gaia \citep[see][for a summary]{2011AJ....142..192F}. The calibration of the CHP distance scale is based on mid--infrared observations of Cepheids in the Milky Way and Local Group galaxies. The distances to individual galaxies are measured by comparing the period--luminosity (PL) relations of their Cepheid populations to that of the Milky Way. In this paper we establish a precise and accurate distance to the Local Group dwarf galaxy IC~1613 using its known population of Cepheids. 

Although the slope of the PL relation at a given wavelength is not observed to vary from galaxy to galaxy \citep{2006ApJ...642..216P}, the sensitivity of the zero--point to various factors is still being debated \citep[e.g.][]{2011A&A...534A..95S}. For example, there has been much discussion over the last few decades regarding the sensitivity of the PL zero--point to metallicity \citep[see][for an overview of recent measurements]{2008A&A...488..731R}. To test for such an effect we can either observe a galaxy with a known metallicity gradient (as first suggested by \citealt{1990ApJ...365..186F} using M31, and later applied to M101 by \citealt{1998ApJ...498..181K} and M33 by \citealt{2009MNRAS.396.1287S}), or look at a selection of galaxies of different metallicities with an independent distance indicator \citep[e.g.][]{1993ApJ...417..553L, 2004ApJ...608...42S}.

%ideal for confirming
With $[Fe/H] \simeq -1$ \citep{2001ApJ...550..554D} IC~1613 is more metal--poor than the SMC, making it extremely useful in setting the the low--metallicity calibration of the PL relation. In this paper we compare the distance obtained from the PL relation in the near- and mid-infrared with that measured from the tip of the red giant branch (TRGB) method. Theory suggests that the effect of metallicity in the mid--infrared will be negligible \citep[e.g.][]{1982ApJ...257L..33M}; however, this has yet to be demonstrated conclusively. The uncertainty in the effect of metallicity on Cepheid magnitudes is one of the dominant systematics still remaining in the extragalactic distance scale. The  present test with IC~1613, along with the other metallicity tests described in \citet{2011AJ....142..192F} allow us to measure the size and sense of the effect.
%, resolving one of the last remaining systematic errors limiting an accurate measurement of the Hubble constant. 

IC~1613 was discovered by \citet{wolf_1906}. It is a type IB(s)m dwarf irregular galaxy in the Local Group \citep{1971ApJ...166...13S, 1991trcb.book.....D}, situated at high galactic latitude at a consensus distance of $736 \pm 49 $~kpc\footnote{Average distance based on 57 measurements in NED: \url{http://ned.ipac.caltech.edu/cgi-bin/nDistance?name=IC+1613}}. This converts to a true distance modulus of $\mu = 24.33 \pm 0.07$~mag, which is slightly closer than M31. IC~1613 is highly resolved and its position above the plane of the Milky Way results in low foreground extinction. As a dwarf galaxy, IC~1613 should have low internal extinction (see below), it is an ideal system for which to measure a distance, as well as to test and compare different distance indicators.

Studies of the Cepheids in IC~1613 began with \citet{1963esag.book.....B}, and were followed up by \citet{1971ApJ...166...13S}, who completed the work begun by Baade over forty years before. Baade chose to observe IC~1613 for the same reasons as we do today: a resolved stellar population and low internal extinction. (Baade deduced the latter from the fact that many background galaxies were visible through the main body of IC~1613.) The low extinction meant that any dispersion in the period--luminosity relation would most probably be due to effects intrinsic to the Cepheids themselves (i.e. temperature), rather than differential reddening, say. Sandage (and Baade) found an anomalously and significantly shallower slope of the period--luminosity (PL) relation ($-1.52$ versus $-2.85$ in the blue, the latter established for Local Group galaxies). However, the observations were compromised by calibration issues and dominated by small--number statistics. Later work by \citet{1988ApJ...326..691F} resolved the issue, showing that the slope of the PL relation in the visible did not change between galaxies.

Ours is not the first test of the Cepheid metallicity effect using IC~1613. \citet{1993ApJ...417..553L} compared TRGB distances of several local group galaxies to those from Cepheids and RR Lyrae stars and found no significant trend in $\Delta \mu$ with $[Fe/H]$ in the $I$ band. \citet{2001AcA....51..221U} found 134 Cepheids during the OGLE survey of this galaxy. They compared the $V$, $I$ and Wesenheit parameter $W_{VI}$ Cepheid distances to the TRGB distance and again found no metallicity effect at those wavelengths. We are repeating their test much further redwards with FourStar data in the near--infrared ($J$, $H$ and $K_{S}$) and IRAC data in the mid--infrared (3.6 and 4.5 $\mu$m). The dispersion of the Leavitt law at this wavelength is minimized, not just by the reduction in differential reddening, but because the amplitudes of the Cepheid light curves and also the width of the instability strip are minimized \citep{2012ApJ...744..132M}.

A preliminary measure of the mid--infrared PL relation in IC~1613 was made by \citet[][hereafter F09]{2009ApJ...695..996F} who searched for Cepheids in archival \textit{Spitzer} data. They found six Cepheids in the cold--mission data and presented PL relations in the 3.6 and 4.5 $\mu$m bands. We compare our results to previous work in Section~\ref{sec:independent_comparison}.

\section{Observations, Data Reduction, and Calibration}
\label{sec:observations}

\subsection{FourStar: $J, H$, and $K_s$}
\label{sec:fourstar_obs}

IC~1613 was observed on three nights using the recently commissioned FourStar wide--field, near--infrared camera \citep{fourstar} on the Magellan Baade 6.5~m telescope at Las Campanas. Table~\ref{tab:fourstar_obs} contains the dates and exposure details.  IC~1613 virtually fills the 10.8 $\times$ 10.8 arcmin field of view so the sky background was determined by imaging an adjacent (sparse) field before and after each IC~1613 dither sequence.  The same individual exposure times were used but a different number of co-adds and dithers were used to save on overhead.  Sources were detected and masked in both the IC~1613 and sky frames.   For each IC~1613 frame the nearest nine (in time) sky frames were combined using the unmasked region in common, then scaled to and subtracted from the on--target frames.  The IC~1613 frames were then combined using an average with sigma-clipping and input rejection masks. This procedure is not expected to produce either flat or sky subtracted frames with zero offset. However, because the stellar photometry uses local sky measurements, the systematic errors due to offset levels should be negligible.

The photometry of the FourStar data was performed with \daophot~\citep{1987PASP...99..191S}. For each field we identified stars to a signal-to-noise ratio $S/N \sim 3$.  We generated a model of the PSF across the field from 100 isolated bright stars.  We then performed PSF fitting photometry with \allstar~allowing the PSF model to vary linearly with $x$ and $y$ across the field. 

To determine the precision of the photometry we added 650,000 artificial stars across each mosaic image. Stars were laid down in a grid with spatial offsets between stars of 40 pixels (10,000 stars at a time) so as to not increase the crowding. The process was repeated with different grid locations until the library of artificial stars was accumulated. We recovered the positions and photometry of the artificial stars by rerunning the same \daophot~routines used for the actual photometry. We achieved better than 10 \% precision for stars brighter than $J\sim21.0$, $H\sim19.5$, and $K_s\sim18.5$ mag.  At the high luminosity end the precision is better than 3\%.

Photometric zero--points were determined for each epoch by matching 18 bright, but unsaturated, stars in the field of IC~1613 to the Two Micron All Sky Survey \citep{2006AJ....131.1163S}. The uncertainties in the zero--points were found to be $\pm 0.015$~mag or better.  However, the scatter around the zero-point for the third epoch of data was about 4 times larger than for the first two epochs.  The larger third epoch scatter was determined to have been caused by excellent seeing. Stellar profiles were $\sim 0.3 \arcsec$ FWHM, causing the PSF to be under-sampled. We applied a smoothing kernel of 1.5 pixels across the mosaics and reran the photometry. This reduced the scatter between 2MASS and FourStar for that epoch to levels comparable to the first epoch. It also reduced the scatter between the FourStar photometry across all three epochs. It did not, however, change the zero--point significantly, viz., less than 0.005 mag.

\subsection{Spitzer IRAC: 3.6 and 4.5~$\mu$m}
\label{sec:irac_obs}

The mid--infrared observations presented here were taken as part of the Warm Spitzer Program PID 61001. We observed in both the 3.6~$\mu$m and 4.5~$\mu$m channels with a frame time of 30~s. The galaxy was observed twelve times over fifteen months between 2010~January~26 and 2011~March~5. The observations were split into three blocks of roughly one month each and were spaced approximately evenly over that time. The dates of the observations are given in Table~\ref{tab:observations}.

The IRAC camera on \textit{Spitzer} has two operating channels: the 3.6~$\mu$m detector observes one field and the 4.5~$\mu$m detector simultaneously observes a closely adjacent field. At any given time one of the detectors will be centered on the target position and the other will be offset. When IRAC takes an exposure it can record both fields simultaneously. This is an advantage in programs such as ours, where a whole galaxy is to be surveyed, as it can cut down the total observation time needed to cover the field in both bands.

However, as the year progresses, \textit{Spitzer} rotates about its axis so the position of the off--target field rotates around the on--target field. This means that in two thirds of our observations, the 3.6~$\mu$m off--target observations are to the SE of the main field and in one third they are to the NW (and vice versa for the 4.5~$\mu$m observations). Therefore, any objects that were not in the ``main" field were imaged in only four or eight epochs, rather than all twelve.  Note however, that all of our detected Cepheids are in the portion of the image covered by all twelve epochs.

Each observation produced two offset maps, containing the galaxy and its surrounding area, covering approximately $0.15 \times 0.10$ degrees, with a subset roughly $0.10 \times 0.10$ degrees covered by both channels. 

\subsubsection{IRAC Mosaic Creation}
\label{sec:mosaics}

The data were analyzed in two ways: a time--resolved analysis using one mosaic per channel per epoch of observation, and an averaged analysis using a single mosaic per channel, comprising all of the time domain data for that wavelength. In addition to these mosaics, a ``master" mosaic was built from all the frames in both channels (1944 frames) and used to determine accurate positions for all the stars in the field. The procedures are described in Section~\ref{sec:allframe}.

\subsubsection{Time--averaged mosaics}
\label{sec:averaged_mosaics}

\begin{figure}
\begin{center}
\includegraphics[width=170mm]{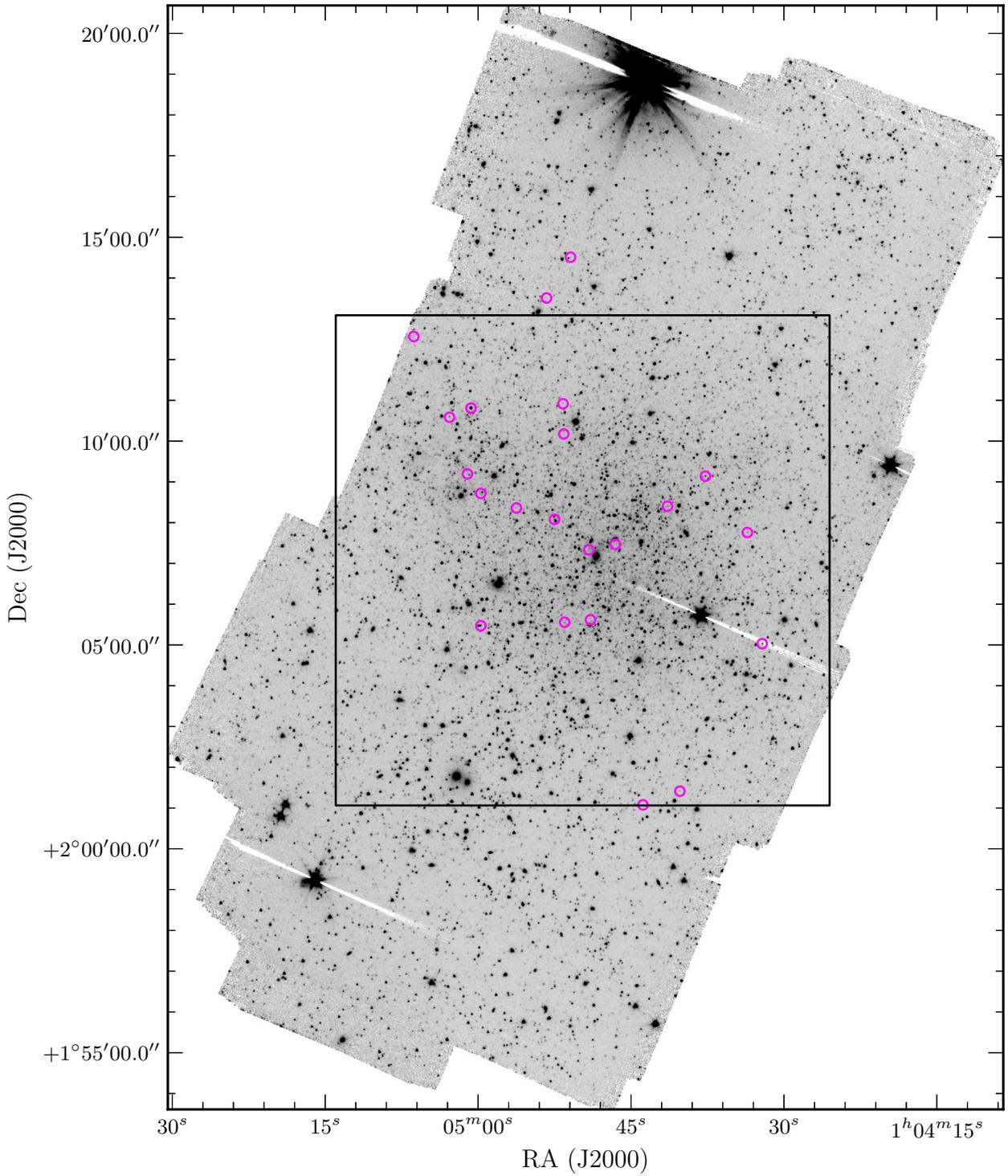}
\caption{IRAC 3.6~$\mu$m science mosaic. The central third is covered in all epochs. The black box shows the region observed with FourStar. Magenta circles denote the positions of the Cepheids. Orientation: North is up, East is left. }
\label{fig:science_mosaic}
\end{center}
\end{figure}

The data from all epochs were stacked into two mosaics (one per channel) using \mopex~\citep{2005ASPC..347...81M}. Each mosaic comprises 960 individual basic calibrated data frames (BCDs). Because of the rotation of the telescope, as described in Section~\ref{sec:observations}, the exposure time coverage within the mosaic is not uniform. The region containing the galaxy has an average coverage of approximately 45 minutes per pixel in each channel, and has three times the coverage of the cold mission data described in F09. These images will be referred to as the ``science mosaics". The 3.6~$\mu$m science mosaic is shown in Figure~\ref{fig:science_mosaic}; the 4.5~$\mu$m mosaic covers the same area.  Approximately the central third of the image has the full twelve--epoch coverage, with the outer thirds having coverage at either four or eight epochs.

Due to the large (1.2") pixel scale of IRAC, stellar profiles are badly under--sampled in single IRAC frames. However, making use of the large number of observations at each spatial position (960) allowed us to resample the images to achieve higher resolution.  The mosaics were created using a pixel scale of 0.75". Other resolutions were tested but 0.75" pixels delivered the best sampled PSF and thus smooth, well--sampled profiles for the stars.

Finally, the science mosaics were converted from MJy~sr$^{-1}$ to data counts using the conversion factors and exposure times in the image headers. The conversion was performed so that \allframe~could give a correct estimation of the magnitude uncertainties.

When creating mosaics, \mopex~preserves the fluxes in the original pixels. This means that the variations in the Cepheid fluxes are not truly lost, they are just averaged over. Although the amplitudes of Cepheids in the mid--infrared can reach 0.6~mag, the average of twelve phase points drawn randomly from the light curve will give a good approximation of the mean flux. For example, if we consider a Cepheid with a mid--infrared amplitude of 0.4 mag (typical of the Cepheids in our study), then the average of 12 random observations will have an uncertainty of 0.03 mag \citep[details of this calculation are in the appendix of][]{2011ApJ...743...76S}. It makes no difference to the Cepheids whether we average these points before or after photometry so long as flux is conserved. However, if we stack the twelve images first then we achieve a higher signal--to--noise ratio image and can detect fainter stars than if we examined single images.  Hence, mosaicking the time--resolved images in a way that preserves flux will give us a good value for the average flux of the Cepheid.

\subsubsection{Single--epoch mosaics}
\label{sec:single_epoch_mosaics}

The single--epoch mosaics were created similarly to the time--averaged mosaics. Each one was made from 81 images and was resampled to a pixel scale of 0.6"\footnote{As the single--epoch mosaics were shallow we did not perform the same tests to find the optimum image scale, instead we used the default value.}. The images were converted from MJy~sr$^{-1}$ to counts using the conversion factors and exposure times in the image headers. Again, the dither pattern meant that the mosaics were not uniformly exposed, but now had typical integration times of 5 minutes per pixel. Note that the single--epoch mosaics are shallower than the data used by F09. Consequently, the time--averaged mosaics were used for the final photometry, while the time--resolved data were used only to confirm that the stars identified as Cepheids were truly variable. 

\subsubsection{Correction mosaics}
\label{sec:correction_mosaics}
In addition to the science mosaics, \mopex~was used to make ``correction" mosaics. These are made identically to the science mosaics, mosaicking the location--dependent correction images provided by the Spitzer Science Center\footnote{Details of the location--dependent photometric correction are given in section 4.5 of the IRAC instrument handbook.} in the same geometrical pattern used for the science mosaics. The correction mosaics are necessary as IRAC is not uniformly sensitive over its entire field of view. The non--uniform coverage depth of our science mosaics can further exacerbate the problem. Inspection of the correction mosaics showed that the residual location--dependent effect had mean values of approximately 2\% in both channels, and would reach as high as 7\% at 3.6 $\mu$m and 10\% at 4.5~$\mu$m in particularly non--uniform regions, if left uncorrected.

\subsubsection{\daophot~and \allframe~reduction}
\label{sec:allframe}

The photometry was performed using the \daophot~and \allframe~packages \citep{1987PASP...99..191S, 1994PASP..106..250S}. The mosaics were run through \daophot~to detect the stars and create a point spread function (PSF) model. The detected objects were subtracted and the resulting frame processed again to detect the remaining objects. The PSF model was created using $\sim$ 100 stars in each mosaic. The final photometry was done using \allframe. \allframe~is preferred over \allstar~in studies where there are multiple frames of the same field as it will produce a master detection list. When the photometry is done using the positions in this list, rather than remeasuring the positions from each frame, the number of free parameters in the PSF fit are reduced. This significantly reduces the uncertainty in the final magnitude, and allows better de-blending of close sources. A detailed description of the process can be found in \citet{1994PASP..106..250S}. The master detection list was generated by running \daophot~on the master frame. The coordinate transformations between this frame and all the other mosaics were determined, then were input to \allframe~to produce the instrumental photometry. 

Artificial star tests were performed on the IRAC images to test the precision of the photometry. Following the methodology set out by \citet{1988AJ.....96..909S}, {\sc addstar} was used to add 10,000 stars to the 3.6 and 4.5~$\mu$m images, 100 stars at a time so as not to significantly increase the level of crowding. We achieve 10\% precision for stars brighter than 19.2 mag (3.6~$\mu$m) and 19.3 mag (4.5~$\mu$m). We also find that crowding does not significantly affect the photometry --- the median difference between the input and output magnitudes for the artificial stars was less than 0.01 mag in both channels.

 \subsubsection{Calibration}
 \label{sec:calibration}
  
All the \textit{Spitzer} photometry in the Carnegie Hubble Program is set to be on the standard system defined by \citet[hereafter R05]{2005PASP..117..978R}. As we used mosaics for our photometry rather than single images we did not need to apply a pixel--phase correction. Any effects due to pixel phase should be adequately averaged over by the dithering and mosaicking. Placing the instrumental magnitudes on the R05 system was achieved by using the PSF stars in the science mosaics as local standards.

The PSF stars in the flux--units versions of the science mosaics (i.e. after mosaicking but before conversion to counts) were each measured using the {\sc phot} aperture photometry routine in IRAF\footnote{IRAF is distributed by the National Optical Astronomy Observatory, which is operated by the Association of Universities for Research in Astronomy (AURA) under cooperative agreement with the National Science Foundation.}. The \textit{zmag} parameter in {\sc phot} was set for each channel such that the procedure would output calibrated magnitudes. 

Each of the stars was measured in a 3.6" radius aperture and sky annulus from 3.6" to 8.4", corresponding to the standard aperture set of 3, 3, 7 native IRAC pixels. To convert this to the 12" radius (10 native pixels) standard aperture used by R05, the latest warm mission aperture corrections of 1.128 and 1.127 in channels one and two, respectively (Spitzer Science Center, Private Communication, 2012) were applied to the measured fluxes. 

The \allframe~photometry was calibrated by finding the offset between it and the corrected aperture magnitudes described above and correcting each star in the field accordingly. Finally, the location--dependent correction was applied by measuring the pixel values in the correction mosaic at the positions of each star in our catalog and multiplying the flux of each star by this value. 

\section{The Cepheid Population of IC 1613}
\label{sec:cepheid_population}

Cepheids were identified by matching to the catalog from the OGLE study \citep{2001AcA....51..221U}. We also identified V22 from \citet{1971ApJ...166...13S}, a long--period Cepheid which was not included in the OGLE catalog, but was included by F09. The Cepheids were initially identified by their positions derived from the science mosaics. The science mosaics were then visually inspected to check for possible nearby contaminants. The light curves generated from the single--epoch mosaics were inspected to check for variability, but this became increasingly difficult at periods below ten days as the uncertainties on the individual points were comparable to the amplitudes of the light curves.

Thirty-one stars measured at 3.6 and 4.5 $\mu$m were matched with the OGLE catalog, and of these twenty-two were measured at $J, H$, and $K_s$. This is just under 25\% of the original OGLE sample. Unfortunately, the majority of Cepheids in IC~1613 have periods below 6 days. Short period Cepheids are naturally fainter and the majority of the OGLE sample fell below our detection limit. 

Several of the Cepheids detected in the IRAC images appeared anomalously bright for their known periods. After visual inspection of both the near-- and mid--infrared images the stars were deemed to be blended with unresolved companions and were excluded from all further analysis. After a review of all of the Cepheids, blends were removed and twenty Cepheids remained. The photometry of the final sample of Cepheids is given in Table~\ref{tab:cep_phot}. The near--infrared time series data is given in Tables~\ref{tab:j_phot}, \ref{tab:h_phot} and \ref{tab:k_phot}.

\subsection{Near--Infrared Period--Luminosity Relations}
\label{sec:near_ir_pls}

Period--luminosity relations were obtained in the $J$, $H$, and $K_{s}$ bands. In this case we had three epochs that were reduced individually. The magnitudes are weighted means of the three observations of each Cepheid, and the uncertainties are the errors of the weighted mean. The systematic uncertainties in the near--infrared photometric zero--points are 0.017, 0.020 and 0.021 mag in $J$,$H$, and $K_{s}$, respectively. 

\begin{figure}
\begin{center}
\includegraphics[width=170mm]{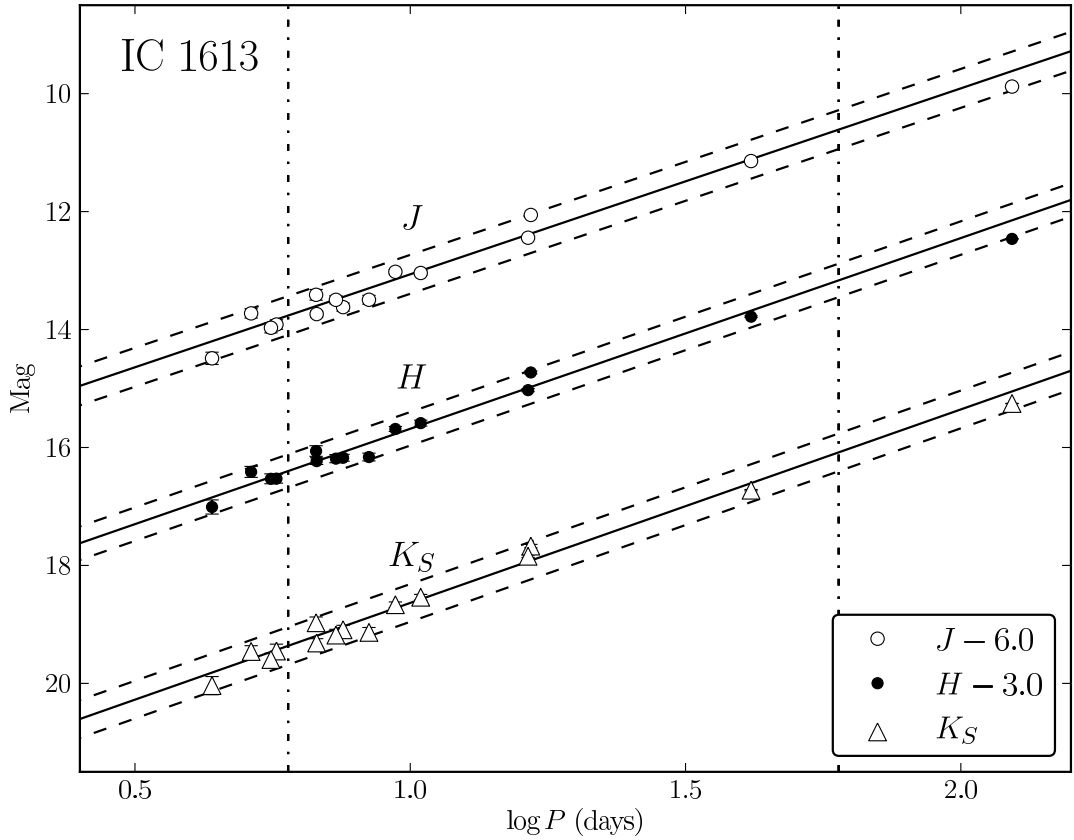}
\caption{Period--luminosity relations in the $J$, $H$ and $K_{s}$ bands. The solid lines represent the fitted PL relations; the dashed lines delineate the +/-2 sigma width of the instability strip. The vertical dot-dash lines show the period range (6 to 60 days) used to fit the PL relations.}
\label{fig:jhk_pl_relations}
\end{center}
\end{figure}

The period--luminosity relations for the near--infrared $J$,$H$, and $K_{s}$ bands are shown in Figure~\ref{fig:jhk_pl_relations}. Prior to fitting, the magnitudes were converted from the 2MASS system to the LCO photometric system, using the transformations described in Section 4.6 of \citet{2001AJ....121.2851C}. The LMC PL relations from Table 6 of \citet{2004AJ....128.2239P} are adopted as fiducial and are rewritten in the form
\begin{equation}
M_{\lambda} = a_{\lambda} (\log P - 1.0) + b_{\lambda} \textrm{.}
\label{eqn:pl_format_nearir}
\end{equation}
We fixed the slopes $a_{\lambda}$ to the LMC values and used an unweighted least squares fit to find the zero--points $b_{\lambda}$. The resulting fits are listed in Table~\ref{tab:pl_relations}. The fits assume $\mu_{0,LMC} = 18.48 \pm 0.03$~mag, as derived in \citet{2012ApJ...759..146M} and \citet{2012ApJ...758...24F}; this puts them on the same scale as the mid--infrared values. The apparent distance moduli derived from the near--infrared PL relations are $24.36 \pm 0.05$, $24.31 \pm 0.04$ and $24.34 \pm 0.05$ mag in $J$, $H$, and $K_{s}$ respectively.

\subsection{Mid--Infrared Period--Luminosity Relations}
\label{sec:mid_ir_pls}

The mid--infrared Leavitt laws take the same form as the near--infrared laws given in Equation~\ref{eqn:pl_format_nearir}, but in this case the slopes were taken from the \citet{2011ApJ...743...76S} LMC results and the zero--points were derived using an unweighted least--squares fit. 

\begin{figure}
\begin{center}
\includegraphics[width=170mm]{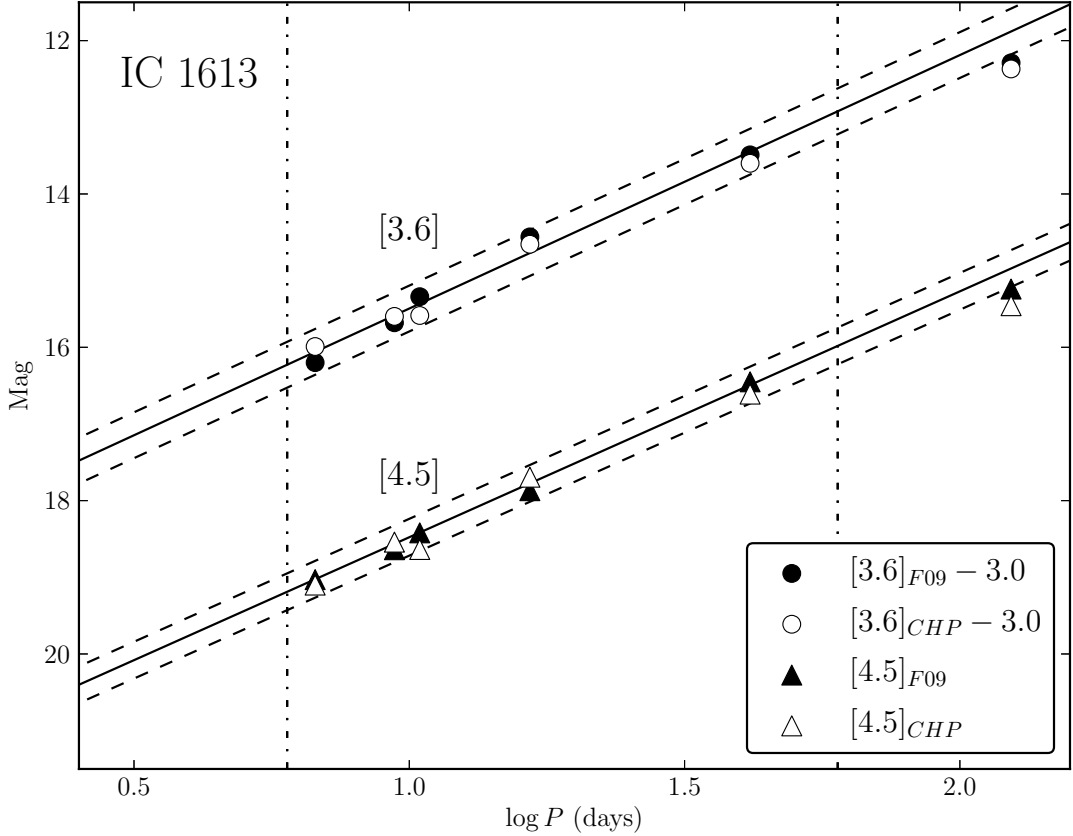}
\caption{Comparison of photometry from this paper with that of \citet{2009ApJ...695..996F} for the six Cepheids they detected. The magnitudes are not expected to be identical as theirs are single--epoch observations, while ours are averages over twelve phase points. The filled symbols are the cold mission data, the open symbols are the CHP data. Solid lines depict the PL fit to the CHP data (fixing the slopes to LMC values, using the sample with $6 \leq P \leq 60$ days) and are indistinguishable from the fits to the cold data. Dashed lines are $\pm2\sigma$ around the fits. The vertical dot-dash lines show the period range (6 to 60 days) used to fit the PL relations.}
\label{fig:pl_cold_comparison}
\end{center}
\end{figure}

The first comparison we make is with the earlier results of F09. They measured single--phase magnitudes for six IC~1613 Cepheids in data obtained from the \textit{Spitzer} archive. The resulting PL relations are plotted in Figure~\ref{fig:pl_cold_comparison}, along with the CHP time--averaged magnitudes for the same six Cepheids.

We do not expect the values to be exactly the same; the amplitude of each Cepheid's light curve will change its position relative to the ridge line of the PL relation in the F09 values (and to a much lesser extent in the CHP values depending on the dispersion of the phase points throughout the pulsation cycle). Hence it it not necessarily useful to compare the magnitudes of individual stars. However, \textit{on average} the results should agree, such that we should get the same result when we fit the PL relation to either set of data.

The PL was found by fixing the slope to the values derived from the LMC \citep{2011ApJ...743...76S} and making an unweighted least--squares fit to the Cepheids with $6 \leq P \leq 60$ days. This period range was chosen to match the Milky Way and LMC samples used to define the IRAC PL relations. The fits to the F09 and CHP data are indistinguishable when plotted; the zero--points differ by only 0.02 mag, which is much smaller than the formal $1\sigma$ error on the fitted zero--points (0.15 and 0.12 mag in [3.6] and [4.5], respectively). 

\begin{figure}
\begin{center}$
 \begin{array}{c c c} 
  \includegraphics[width=30mm]{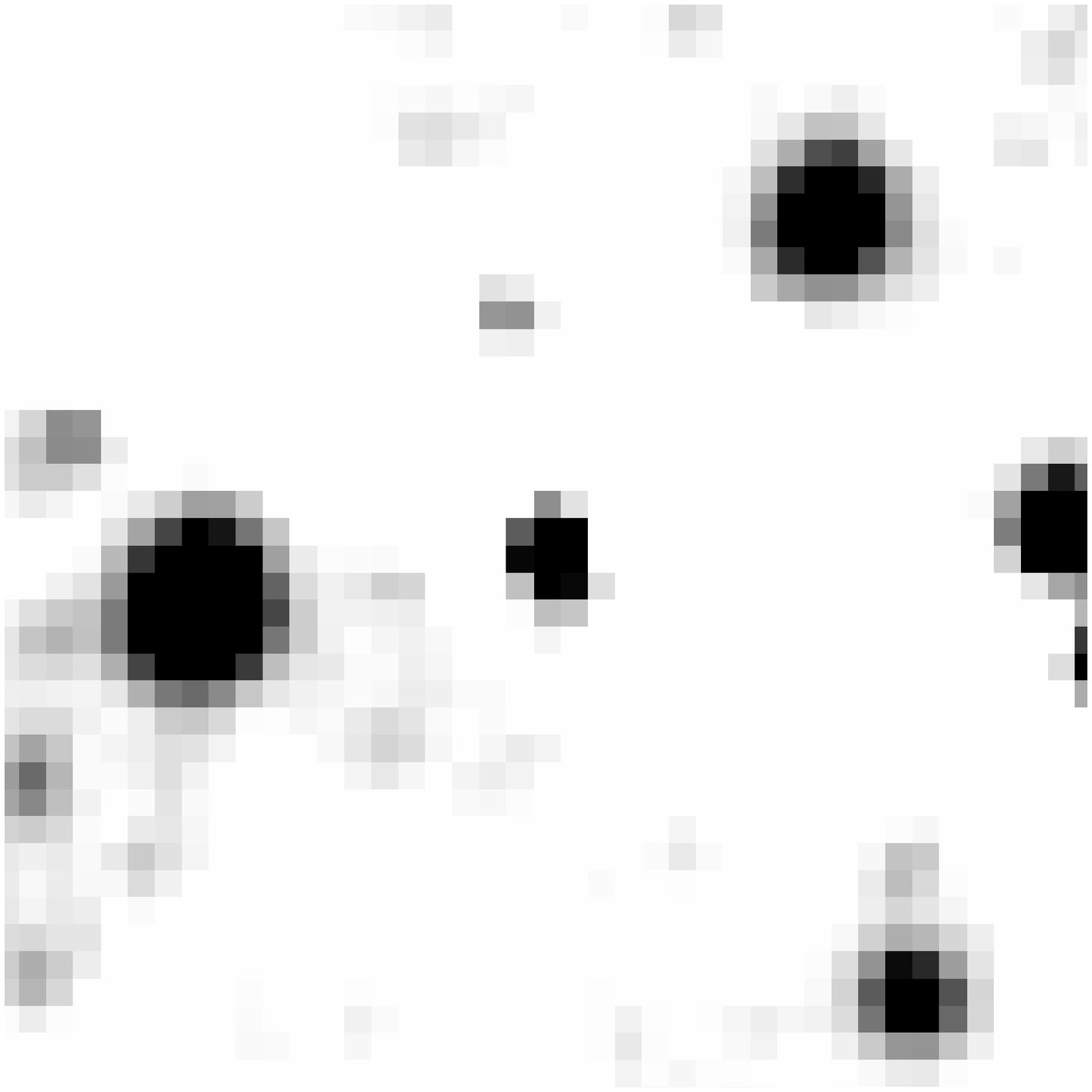} &
  \includegraphics[width=30mm]{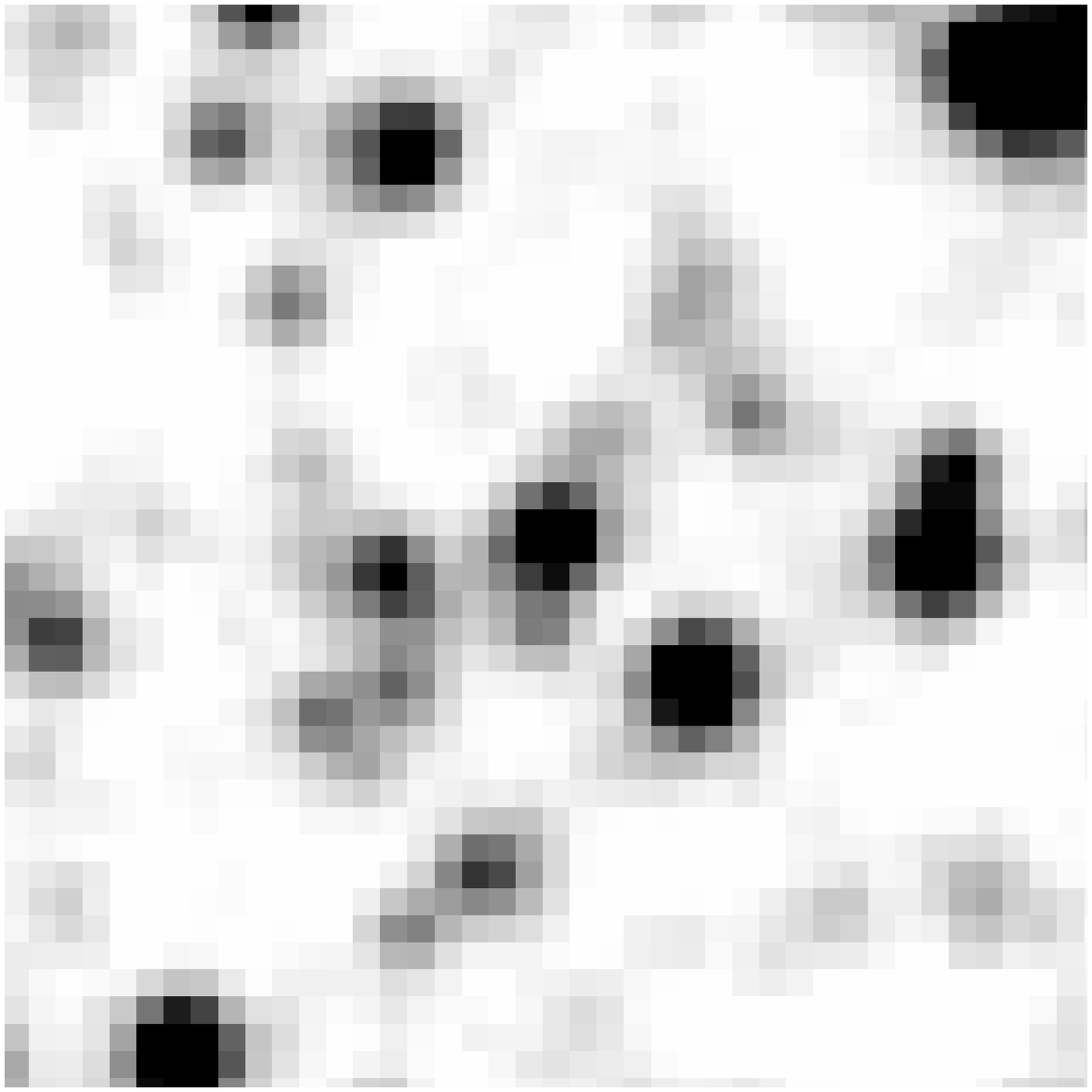} &
  \includegraphics[width=30mm]{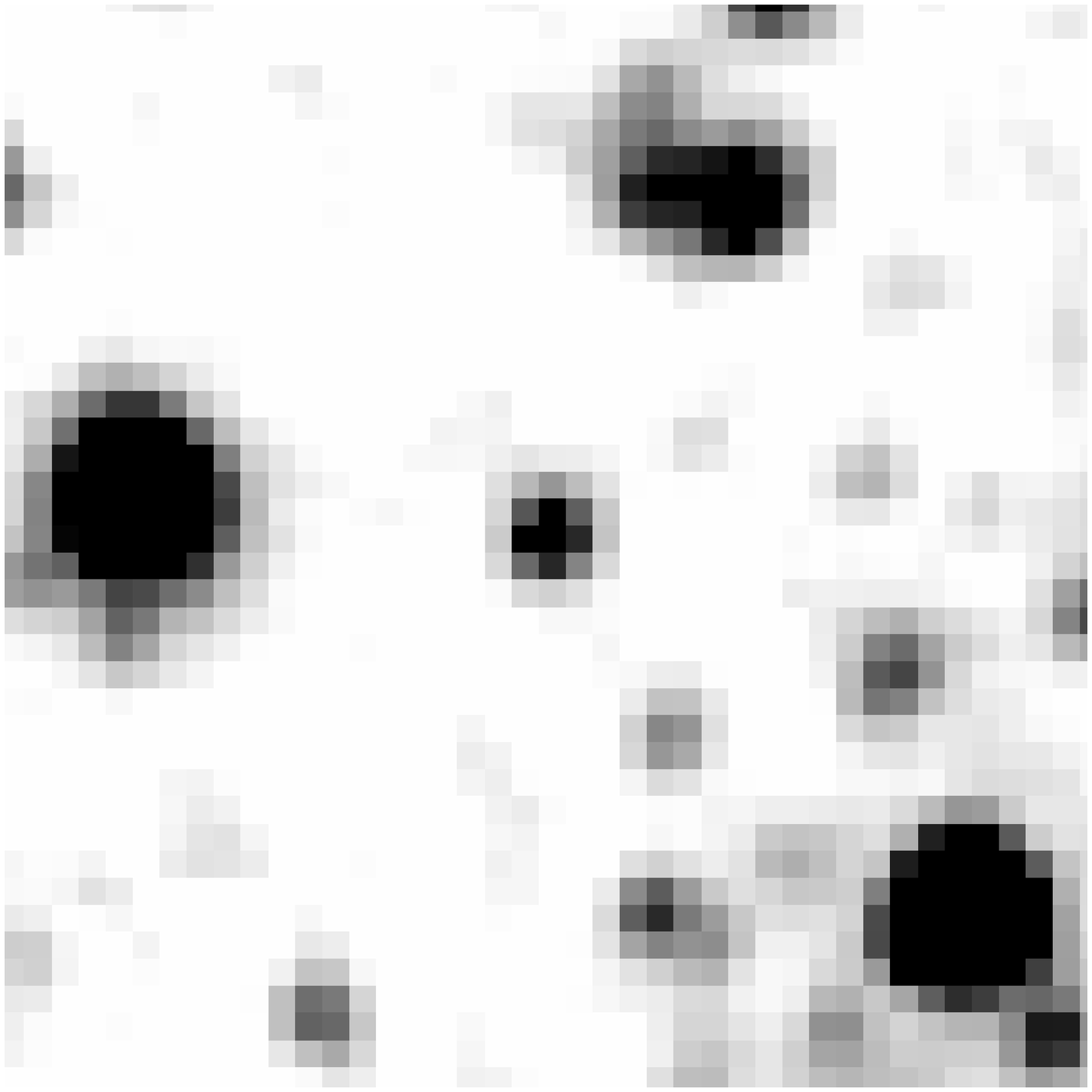} \\
 \mbox{11446} & \mbox{736} & \mbox{7647} \\
  \includegraphics[width=30mm]{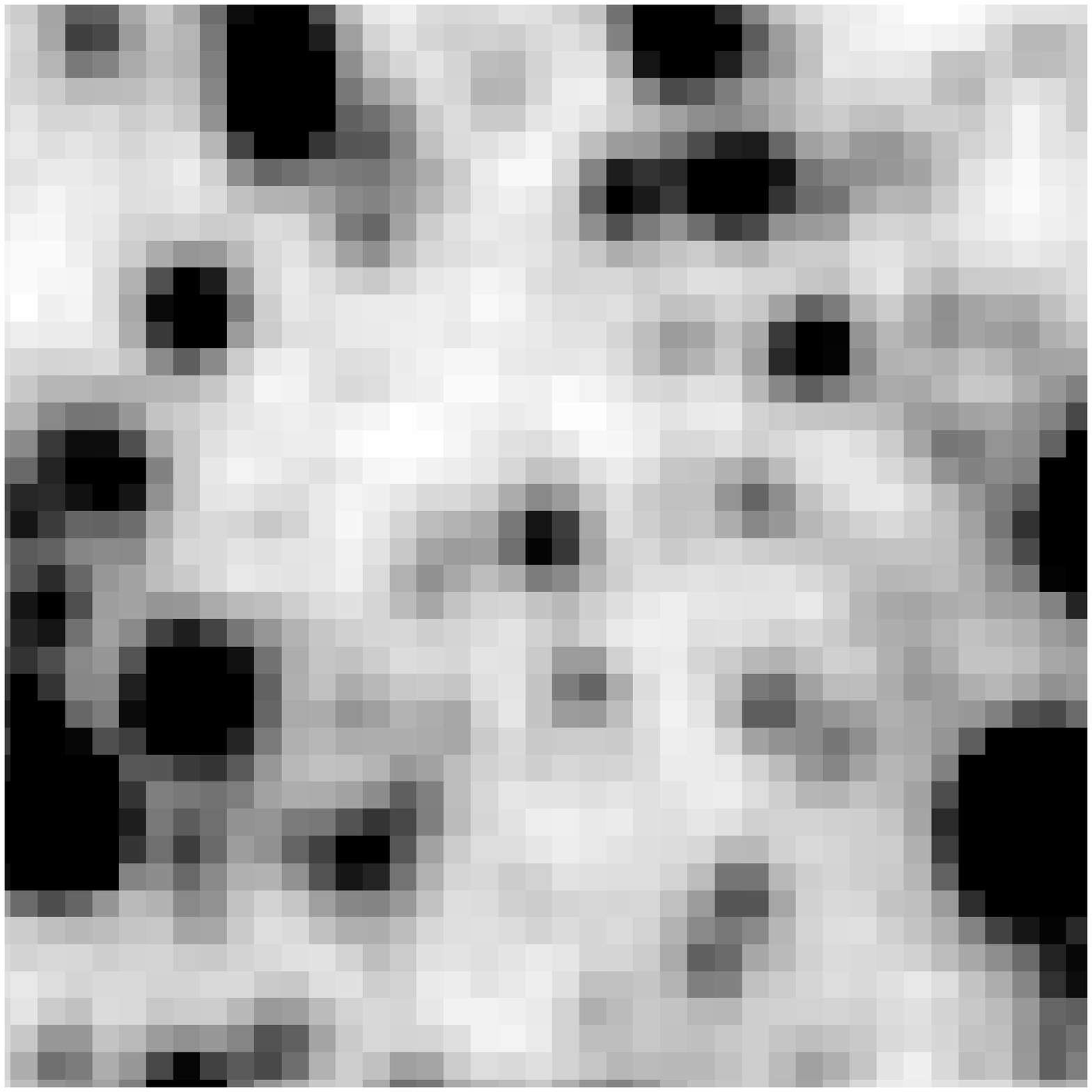} &
  \includegraphics[width=30mm]{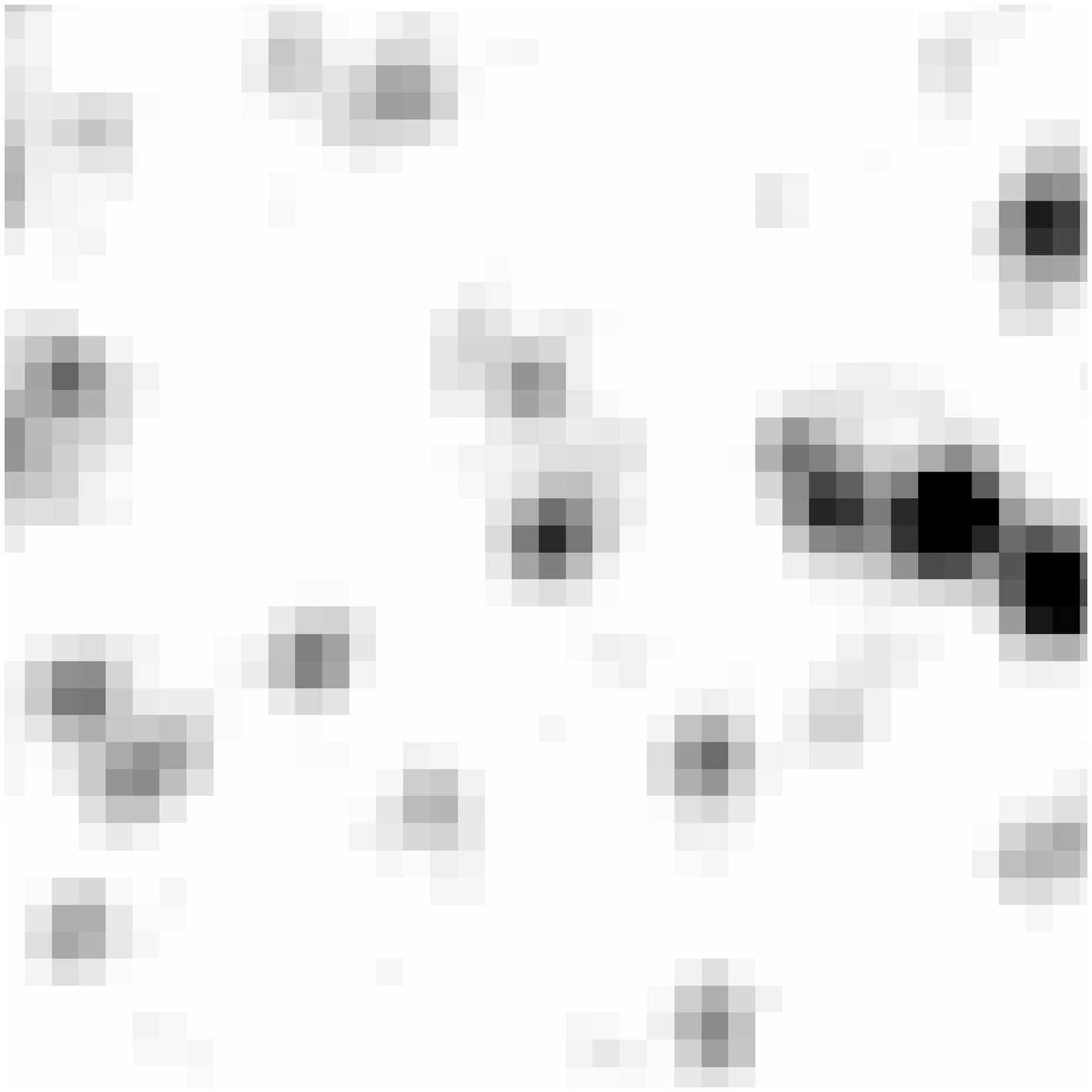} &
  \includegraphics[width=30mm]{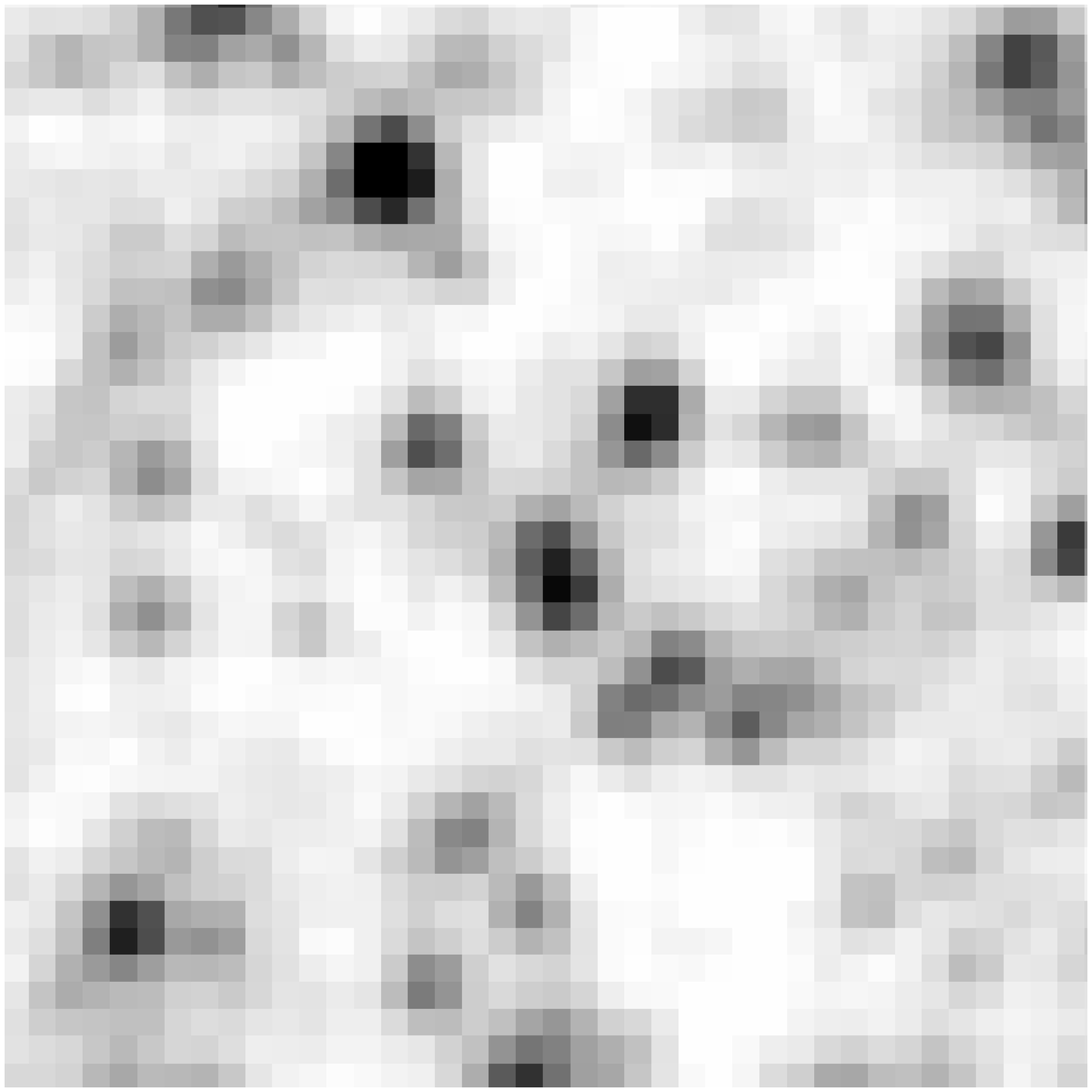} \\
 \mbox{13738} & \mbox{7664} & \mbox{926} \\
  \includegraphics[width=30mm]{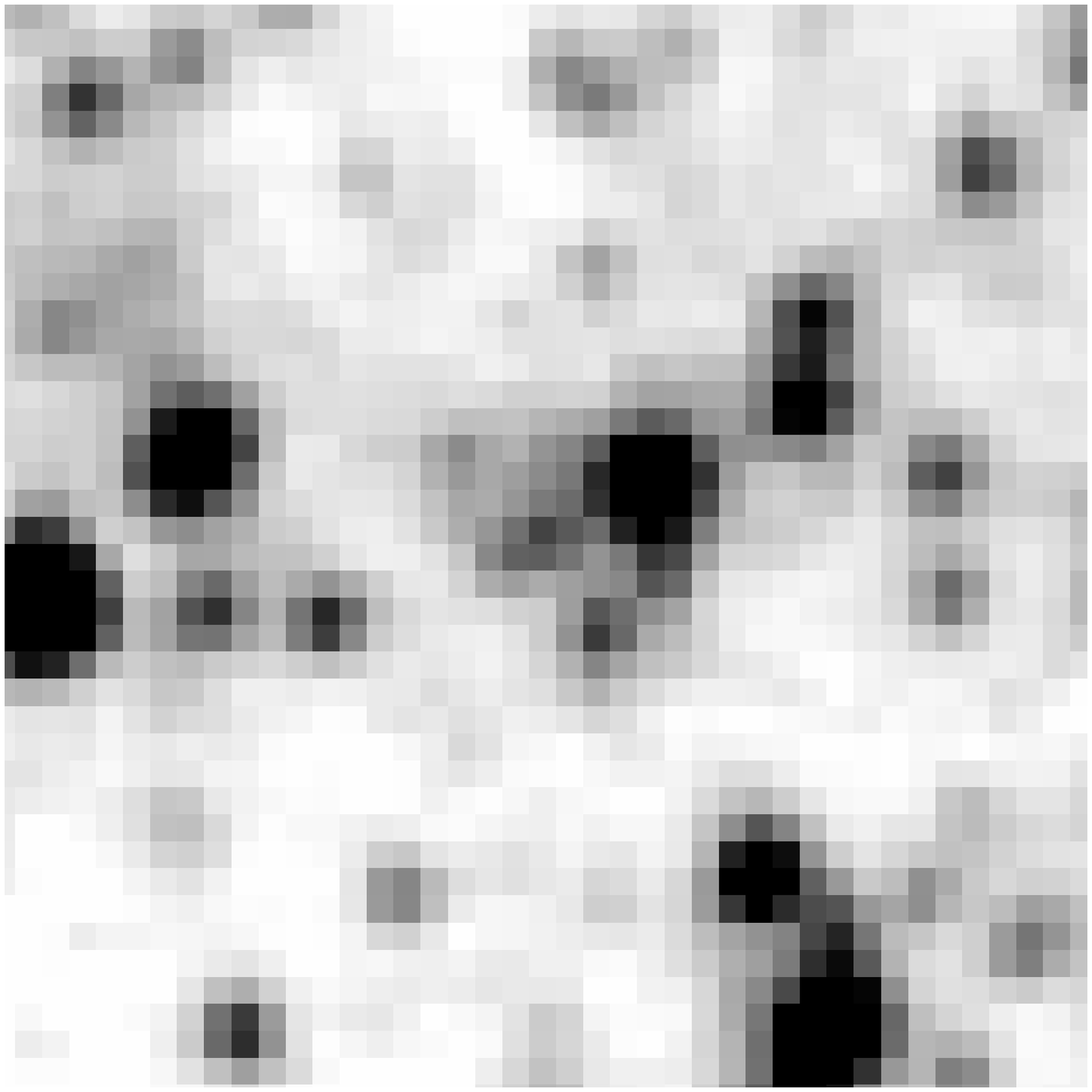} &
  \includegraphics[width=30mm]{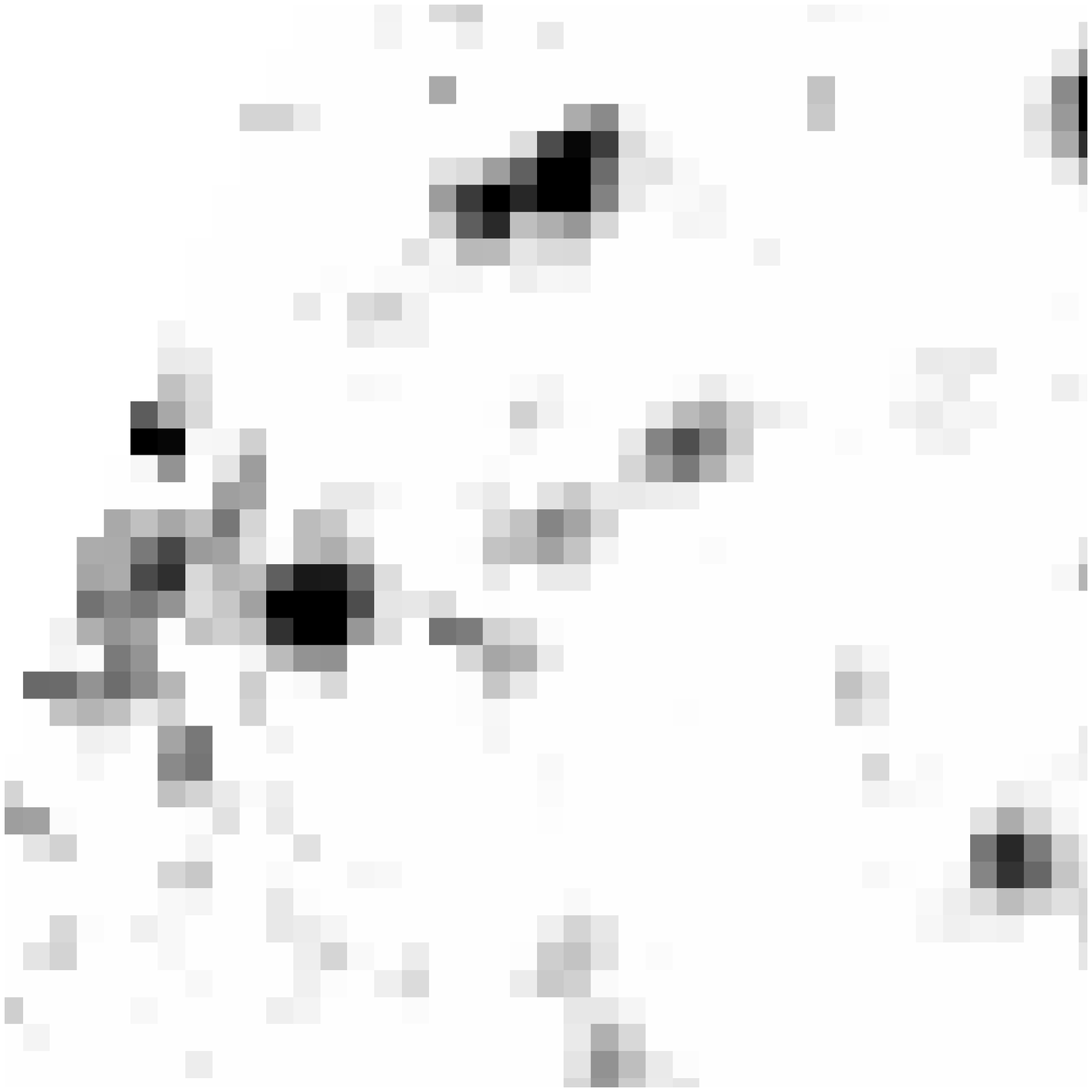} &
  \includegraphics[width=30mm]{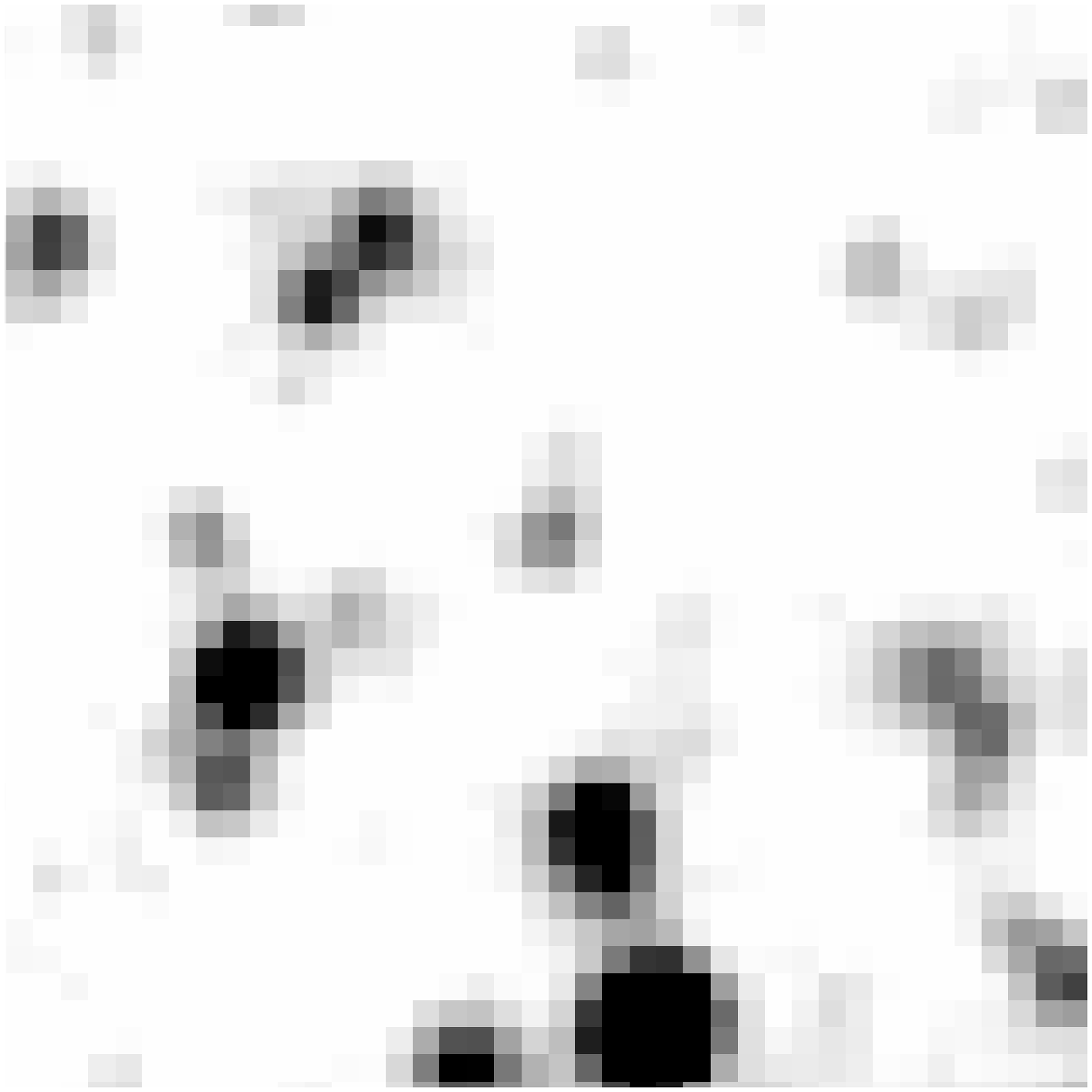} \\
 \mbox{11589} & \mbox{13759} & \mbox{18905} \\
  \includegraphics[width=30mm]{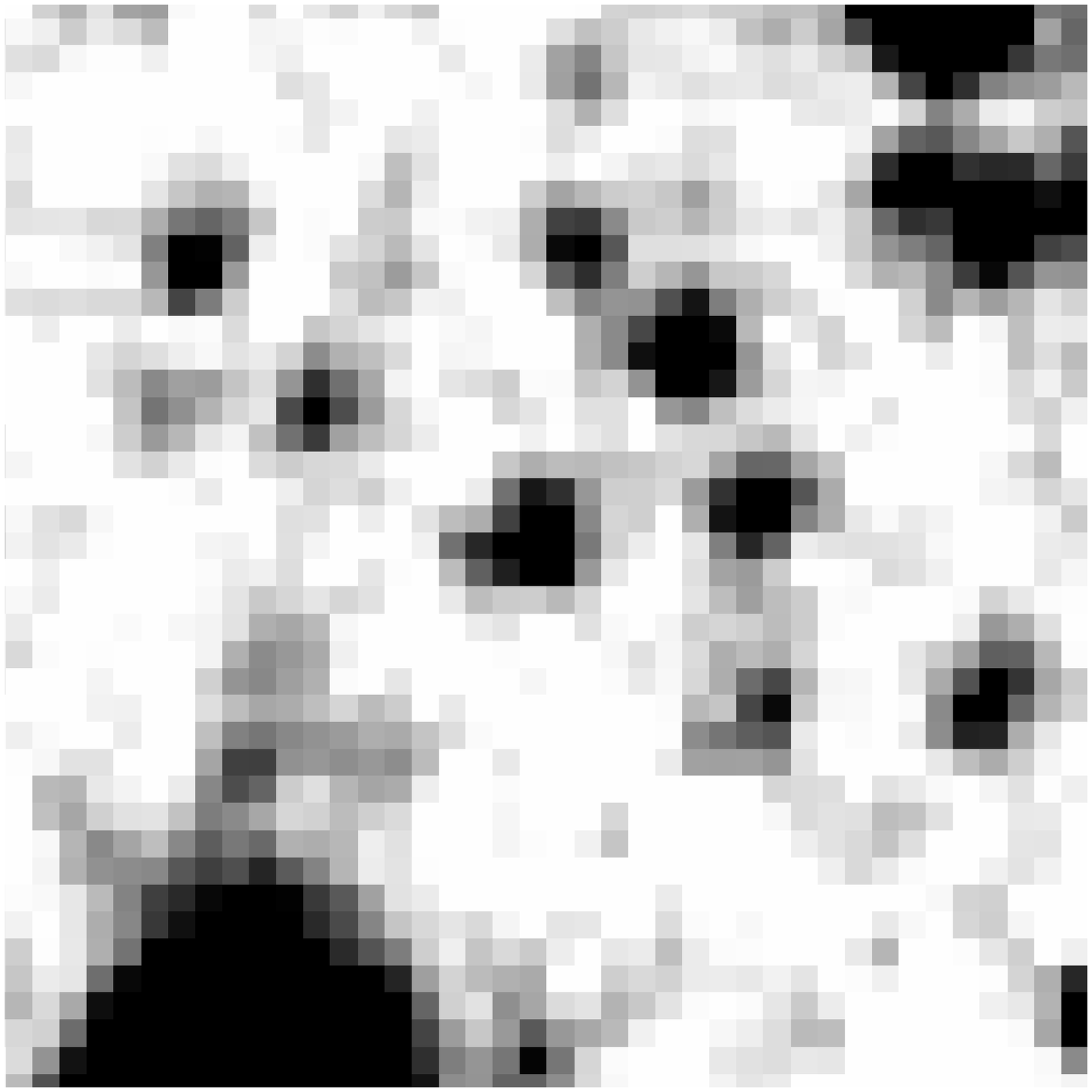} &
  \includegraphics[width=30mm]{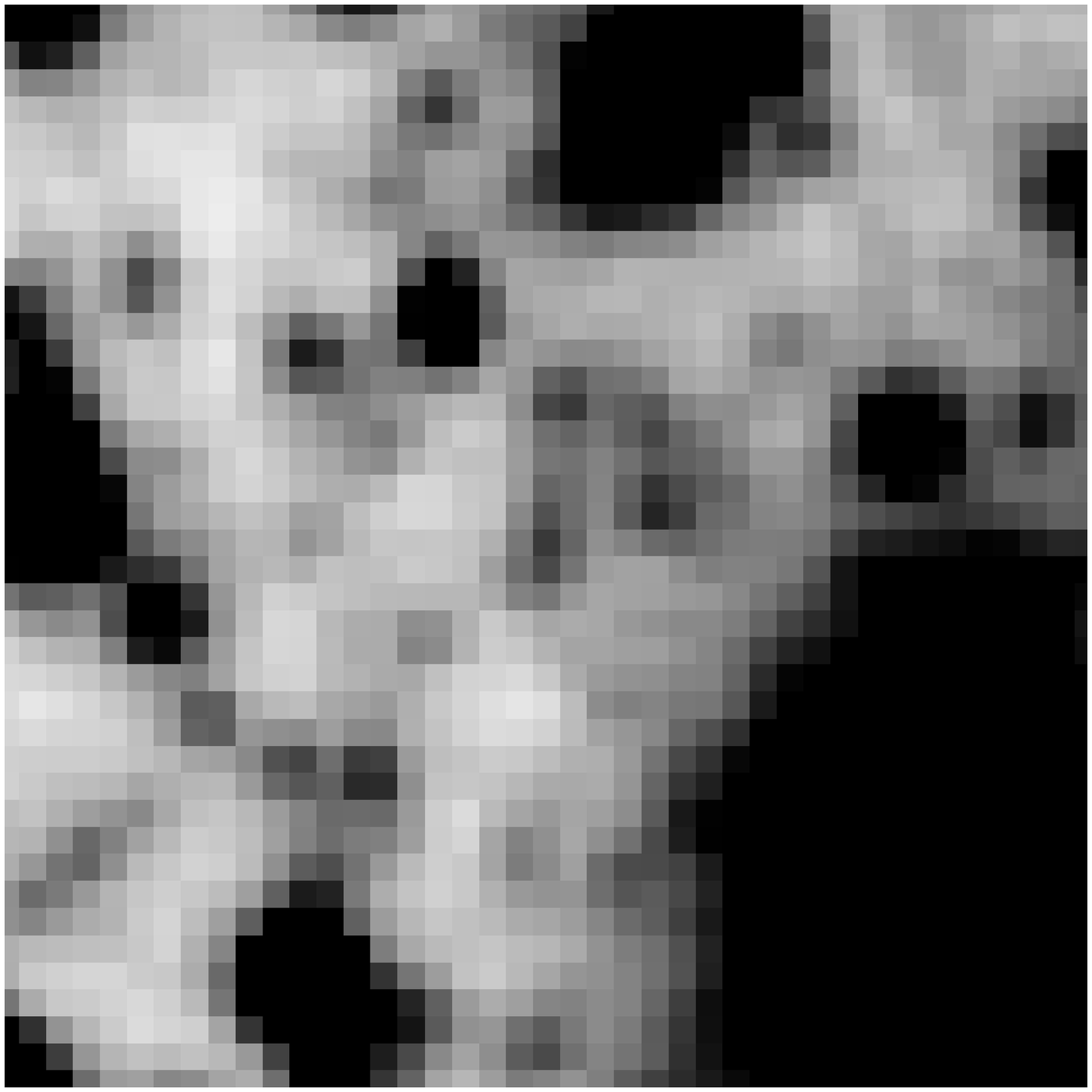} &
  \includegraphics[width=30mm]{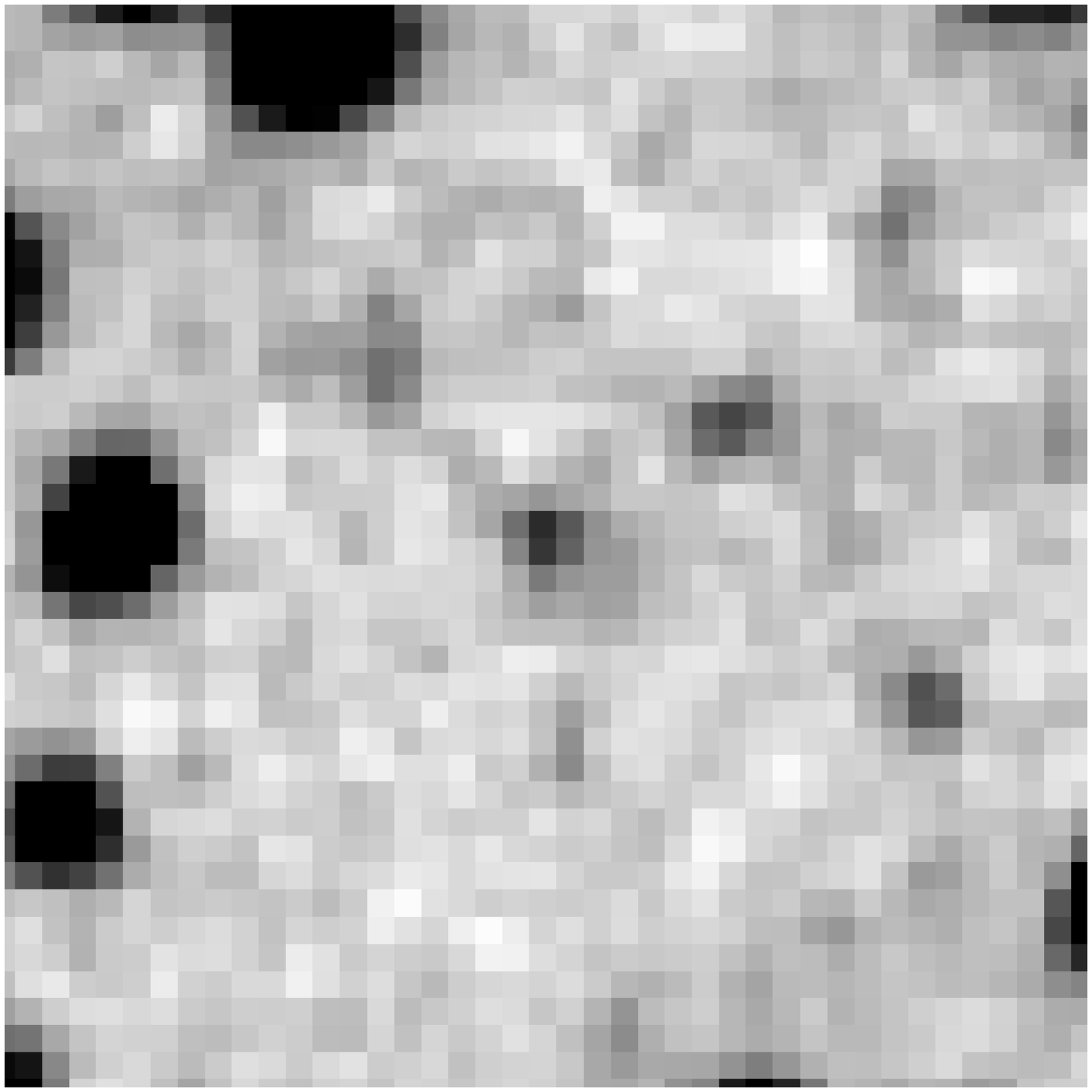} \\
 \mbox{13943} & \mbox{3732} & \mbox{5037} \\

\end{array}$
\caption{A montage of 3.6~$\mu$m images of the Cepheids used in the mid--infrared PL relation. The Cepheids are the stars in the center of each image. The PSF--subtracted image was examined for every Cepheid and the ones in the final PL relation were cleanly subtracted from the image, showing no evidence of crowding.}
\label{fig:cepheid_images}
\end{center}
\end{figure}

\begin{figure}
\begin{center}
\includegraphics[width=170mm]{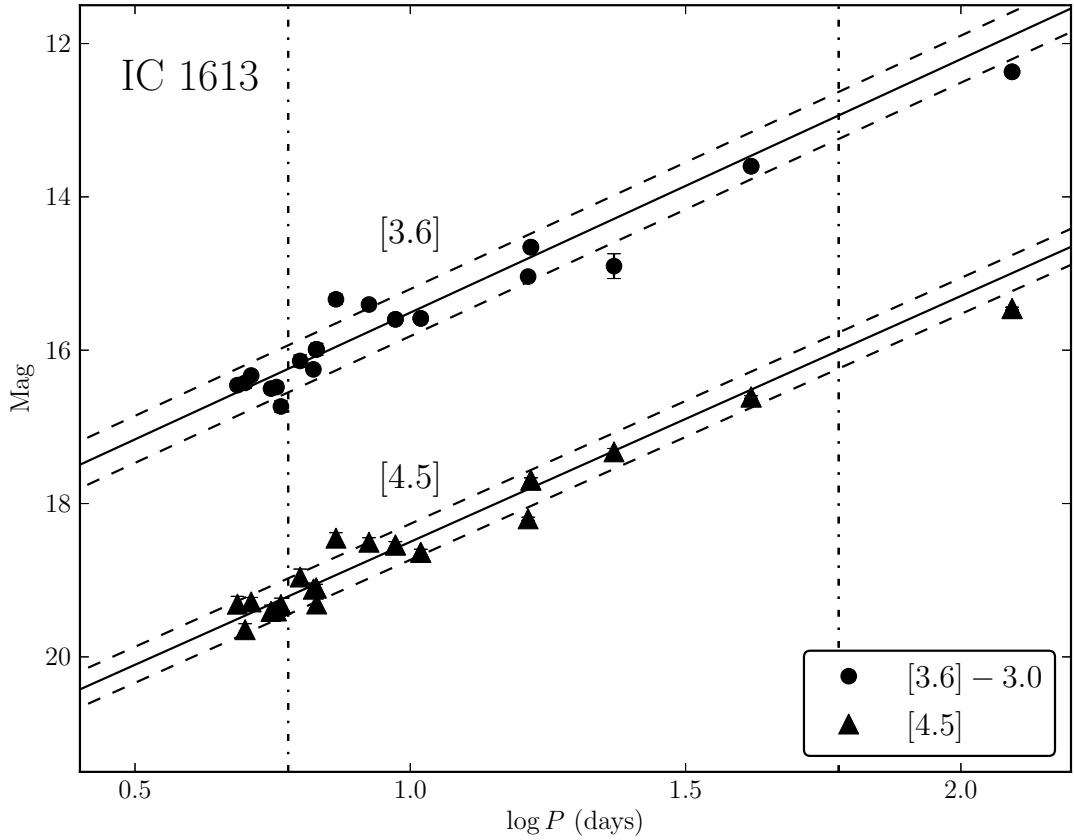}
\caption{Period--luminosity relations in the 3.6 and 4.5$\mu$m bands. The solid lines represent the fitted PL relations; the dashed lines show the $\pm2\sigma$  changes in zero--point. The vertical dot-dash lines show the period range (6 to 60 days) used to fit the PL relations.}
\label{fig:pl_relations}
\end{center}
\end{figure}

Finally, we re--fit the PL using the whole sample of Cepheids in the period range 6 to 60 days. A montage of the 3.6~$\mu$m images of these Cepheids can be found in Figure~\ref{fig:cepheid_images}. The results are given in Table~\ref{tab:pl_relations}, and the relations are plotted in Figure~\ref{fig:pl_relations}. By comparing the calculated zero--points to those in the Milky Way (MW) PL relations given in \citet{2012ApJ...759..146M} we find apparent distance moduli of IC~1613 of $24.31 \pm 0.09$ and $24.26 \pm 0.08$ mag using the [3.6] and [4.5] PL relations, respectively. The quoted uncertainties include the error estimates from both the MW and IC~1613 fits.

\subsection{Reddening Corrected Distance Modulus}
\label{sec:reddening}
\begin{figure}
\begin{center}
\includegraphics[width=170mm]{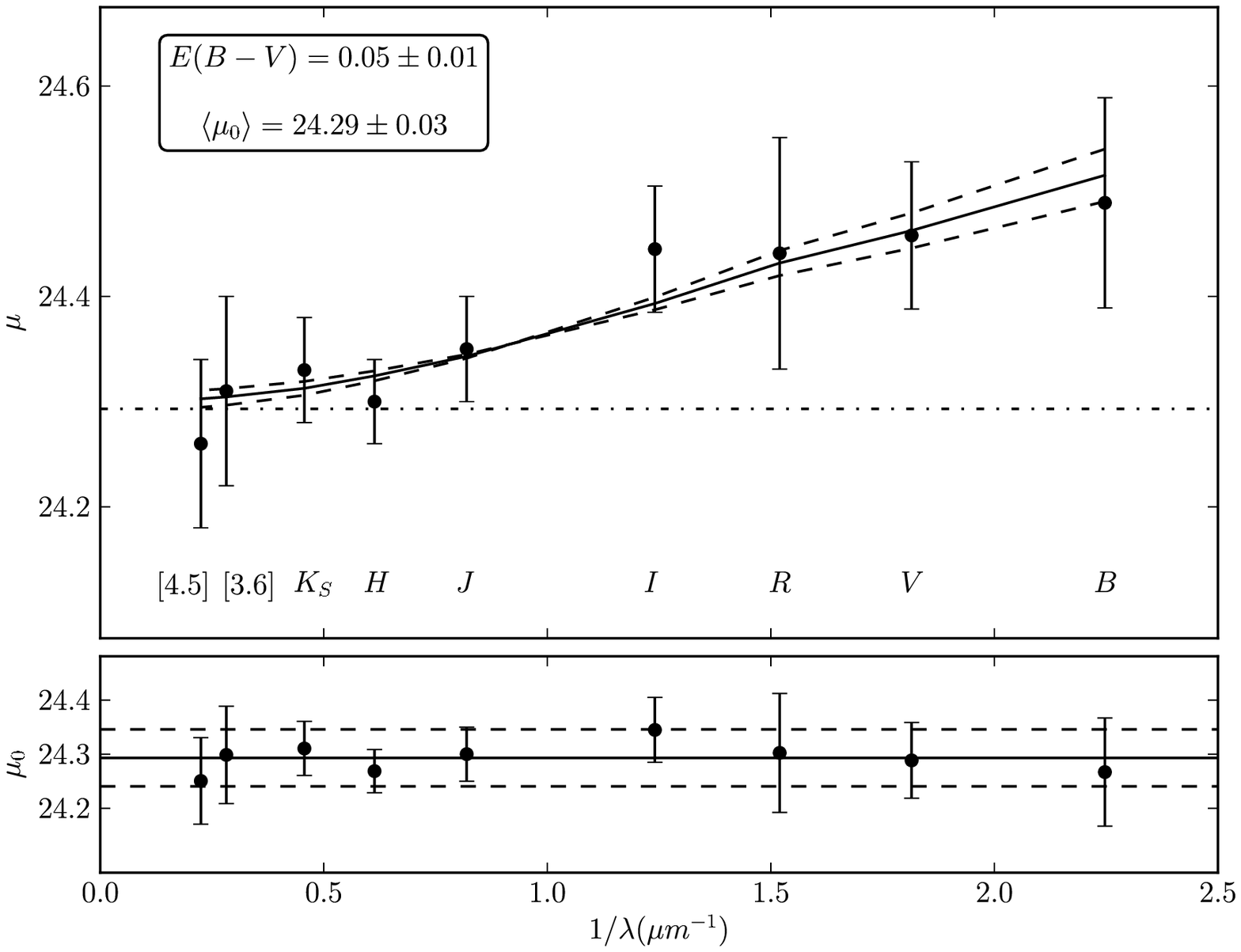}
\caption{Fitting the IC~1613 distance moduli to the reddening laws of \citet{1989ApJ...345..245C} ($B$ to $K_{s}$) and \citet{2005ApJ...619..931I} ($K_{s}$ to $[4.5]$). Points at $B$ and $R$ are taken from \citet{1988ApJ...326..691F}, $V$ and $I$ are from \citet{2001AcA....51..221U}, $J$, $H$ and $K_{s}$ are FourStar data, and $[3.6]$ and $[4.5]$ are IRAC data. The distance moduli from $B$ to $R$ were refit using the LMC PL relations from \citet{2007A&A...476...73F} and adopting $\mu_{LMC} = 18.48$. The solid line is the best--fit reddening law, the dashed lines are $\pm 1 \sigma$ around the law, and the dot-dashed line is the resulting reddening--corrected distance modulus. The bottom panel shows the residuals of the extinction--corrected distance moduli around the mean value.}
\label{fig:reddening}
\end{center}
\end{figure}

We can make use of the broad wavelength coverage of archival observations to derive the total line--of--sight reddening and extinction to the Cepheids in IC~1613. Figure ~\ref{fig:reddening} demonstrates the technique. The distance moduli are plotted as a function of inverse wavelength in microns. Three extinction laws --- the optical and near--infrared laws from \citet{1989ApJ...345..245C} and the mid--infrared law from \citet{2005ApJ...619..931I}\footnote{Use of the \citet{2005ApJ...619..931I} law was justified by \citet{2012ApJ...759..146M}.} --- are combined to fit to the data, assuming $R_{V} = 3.1$. The best fit $E(B-V)$ value was found by minimizing the dispersion of the distance moduli around the scaled and shifted extinction law, and was found to be $E(B-V) = 0.05 \pm 0.01$~mag. The top panel in Figure~\ref{fig:reddening} shows the scaled, shifted extinction law with the apparent distance moduli at nine wavelengths; the dashed lines show how the fit changes if $E(B-V)$ is changed by $1\sigma$. Applying this scaled correction to each of the apparent distance moduli and taking a weighted average results in an absolute distance modulus of $\langle \mu_{0}\rangle = 24.29 \pm 0.03_{stat} \pm 0.03_{sys}$ mag. The deviations of the extinction--corrected distances around this value are shown in the bottom panel of Figure~\ref{fig:reddening}. The systematic uncertainty comes from the LMC distance we adopt: $\mu_{LMC} = 18.48 \pm 0.03$~mag. 

To test the robustness of this technique the analysis was repeated with either one or two data points removed from the input, or with different values for the individual distance moduli. For example, the data was re-fit with either $K_{s}$, $[4.5]$ or both bands removed. None of these solutions was found to affect the resulting reddening correction or distance modulus at a significant level. We also tested the fit by using the original values from Table 4 of \citet{2001AcA....51..221U} (with their reddening correction removed). Despite the fact that the change in zero--point is significant --- over 0.2 mag --- the resulting fit was barely affected. This shows that the nine--band fit is an excellent way to measure the distance modulus to a population, even if the data is somewhat heterogeneous. 

Note that we have made no metallicity correction for the $V$ and $I$ data in this fit. As we will discuss in Section~\ref{sec:independent_comparison}, there is some evidence that the zero--points of the optical Leavitt laws are affected by metallicity and there have been many calculations of $\gamma_{W}$, which measures the change in the optical Wesenheit Leavitt law zero--point with changing metallicity. However, $\gamma$ is much more difficult to estimate for the individual optical bands as metallicity effects are degenerate with reddening. 

We do not believe that the metallicity effect has a significant effect on the value of $E(B-V)$ derived in this section. The majority of the power in the fit is in the long wavelength range where we have more data, which is where the metallicity effect is believed to be vastly reduced. The small changes induced in $V$ and $I$ due to metallicity effects are much less significant than the corrections we have made to put all the data on the same zero--point.

\section{Independent Distance Comparisons}
\label{sec:independent_comparison}

The nine--band fit presented in Figure~\ref{fig:reddening} produces a reddening--corrected Cepheid distance modulus of $24.29 \pm 0.03_{stat} \pm 0.03_{sys}$ mag. In this section we compare our result to other recent measurements also using Cepheids, and then compared to other independent distance indicators.

\subsection{Cepheid Comparison}
\label{sec:cepheid_comparisons}
The Auracaria project is using near--infrared observations of Cepheid populations to determine distances to nearby galaxies. We compare our results with their study of IC~1613 \citep{2006ApJ...642..216P}, in which the authors use the template fitting method of \citet{2005PASP..117..823S} to obtain mean--light magnitudes of Cepheids in $J$ and $K$ from single--epoch observations. They observed 39 Cepheids in the galaxy, the majority of which are also observed in our study. Adopting the PL slopes from \citet{2004AJ....128.2239P} and adopting an LMC distance modulus of $18.50 \pm 0.10$ mag they find $\mu_{J} = 24.385 \pm 0.040$ mag and $\mu_{K} = 24.306 \pm 0.045$ mag. Both of these values agree with our FourStar results within the $1\sigma$ uncertainties. 

\citet{2006ApJ...642..216P} combine their $J$ and $K$ distance moduli with the $V$ and $I$ values derived in OGLE II by \citet{2001AcA....51..221U} to derive a multi--wavelength fit for $E(B-V)$ and the extinction--corrected distance modulus. Using their four--band fit they measure $E(B-V) = 0.090 \pm 0.019$ mag and derive $\mu_{0} = 24.291 \pm 0.035$ mag. Their derived reddening is slightly higher than our value of $0.05 \pm 0.01$ mag, but is barely outside the respective $1\sigma$ error bars. Their de-reddened distance, however, is in complete agreement with our value of $24.29 \pm 0.03$ mag. This goes to show the power of moving to the infrared; the effect of reddening is significantly reduced here such that the choice of reddening law and the value of $E(B-V)$ have little effect at these wavelengths. 

More recently, \citet{2010ApJ...712.1259B} (B10) observed IC~1613 at optical wavelengths using the \textit{Advanced Camera for Surveys} (ACS) on the \textit{Hubble Space Telescope} (\textit{HST}). Their study looked at the fainter variable stars in the galaxy and found 44 Cepheids pulsating in either the fundamental or overtone modes. These Cepheids have short periods (the majority have $\log P < 0.5$) and were not detected in the CHP observations. 

B10 derive the Cepheid distance using the Wesenheit index $W_{VI}$ \citep{1976RGOB..182..153M}. The $W_{VI}$ index is reddening--free by design, hence is only affected by the choice of reddening law, and not at all by the total amount of reddening. Adopting an LMC distance modulus of $18.515 \pm 0.085$ \citep[from][]{2003AJ....125.1309C}, they find $\mu_{0,W} = 24.50 \pm 0.11$ using only the fundamental mode Cepheids. This value is significantly higher than the distance we find in Section~\ref{sec:reddening}. However, it is possible that this is due to the lack of a metallicity correction on $\mu_{0,W}$.

B10 make no correction for metallicity in their Cepheid analysis, and note that they do not believe a correction is necessary at the low metallicity of IC~1613. It may be the case that the metallicity correction required on $\mu_{0,W}$ decreases as we move to low $[Fe/H]$ populations, but this has not been proven conclusively. To this end we take the correction on $\mu_{0,W_{VI}}$ from \citet{2009MNRAS.396.1287S} of $\gamma_{W_{VI}} = -0.29 \pm 0.11$~mag~dex$^{-1}$ and apply it to the distance derived by B10. Assuming $12 + \log(O/H)$ of 8.34 \citep{2004ApJ...608...42S} and 7.90 \citep{2007ApJ...671.2028B} for the LMC and IC~1613 respectively, and now adopting the CHP LMC distance modulus of $18.48$ mag, we find a metallicity corrected distance modulus of $\mu_{0,W,Z} = 24.33 \pm 0.14$ mag. This is now consistent with our value derived in Section~\ref{sec:reddening}, but its error bar is much larger than the original B10 result. 

We consider the same approach for the optical data from \citet{2001AcA....51..221U}. Adopting the CHP LMC distance modulus we recalculate their reddening--free distance modulus to be $\mu_{0,W} = 24.41 \pm 0.07$ mag; again, higher than the reddening--corrected distance derived in this paper. Adopting the same metallicity parameters as the previous paragraph we find a reddening--free, metallicity--corrected $\mu_{0,W,Z} = 24.24 \pm 0.11$ mag. Like the B10 result, this is consistent with our result, but the metallicity correction has driven up the uncertainty on the value.

It is possible that the large difference between the our distance modulus and that from B10 is due to the very different period distributions of the two samples. B10 focusses on short period ($\log P < 0.5$) Cepheids, while this work studies a longer period sample ($0.77 < \log P < 1.77$). The linearity of the Leavitt law in the Wesenheit bands was recently studied bu \citet{2013MNRAS.431.2278G}. They found that the Wesenheit law may have a break around 10~days, and that metallicity may play a part. It is clear that the metallicity effect on the Leavitt Law zero--point in the optical bands requires more study, and that its effect at the lowest metallicities is not yet conclusively ruled out.

\subsection{RR Lyrae Comparison}
\label{sec:rrlyrae_comparison}
The \citet{2010ApJ...712.1259B} study not only found Cepheids in IC~1613 but also RR Lyrae stars. RR Lyrae stars obey a luminosity--metallicity relation at optical wavelengths. Adopting a mean metallicity of [Fe/H]$ = -1.6 \pm 0.2$ and the luminosirt--metallicity relation from \citet{2009ApJ...699.1742B}, they calculate the absolute magnitude of the horizontal branch in IC~1613 to be $M_{V} = +0.52 \pm 0.12$ mag, and derive a reddening corrected RR Lyrae distance modulus of $24.39 \pm 0.12$~mag; but see Section~\ref{sec:dispersion} and the RR Lyrae comparison panel in Figure~\ref{fig:distance_comparison}. This is larger than, but still within $1\sigma$ of our multi--band Cepheid fit presented above.

\subsection{Tip of the Red Giant Branch Comparison}
\label{sec:trgb_comparison}

An independent measure of the distance to IC~1613 can be obtained using the tip of the red giant branch (TRGB). The absolute magnitude of the TRGB is a physical property of the stellar population and does not depend on any measurements further down the distance ladder. 

An excellent review of the use of Color-Magnitude-Diagram based distance indicators, including the TRGB, is \citet{2012Ap&SS.341...65S}. Briefly summarized, the $I$ band TRGB is considered a robust distance indicator because the bolometric correction to the $I$ band magnitude as a function of $[Fe/H]$ and effective temperature is complementary to the changes in bolometric luminosity due to differences in metallicity. These two effects cancel each other in the $I$ band, meaning that the absolute $I$ band magnitude of the TRGB is essentially constant with both age and metallicity; although see \citet{2009ApJ...690..389M} for their T-magnitude technique and calibration of even this small residual metallicity effect.. This makes the TRGB a robust measure of distance for old resolved stellar populations, and an independent check of the Cepheid distance moduli we have presented in the previous section.

%Moreover, in the $I$ band its absolute magnitude is relatively insensitive to metallicity and age effects, making it a robust measure of distance for old resolved stellar populations, and an independent check of the Cepheid distance moduli we have presented in the previous section.

We compare our result to the work of \citet[hereafter D01]{2001ApJ...550..554D}, who derived the TRGB distance to IC~1613 using $V$ and $I$ band photometry from WFPC2 on \textit{Hubble}. They provide two estimates of the TRBG apparent magnitude --- $20.40 \pm 0.09$ from their own data and $20.35 \pm 0.07$ from a re-reduction of the data from \citet{1999AJ....118.1657C}. The second value is more robust as it is measured from a region with higher stellar density, hence more stars on the red giant branch. They assume the absolute magnitude of the TRGB is $M_{I} = -4.02 \pm 0.05$, and a foreground extinction of $A_{I} = 0.05 \pm 0.02$, resulting in an extinction--corrected distance modulus of $\mu_{0} = 24.32 \pm 0.09$ mag.

The true distance modulus derived from our multi--band fit is consistent with the TRGB distance from D01. This suggests that metallicity effects are not significantly affecting the Cepheid distance modulus derived here; our PL relations were all calibrated to the MW and LMC which have much higher average metallicities than IC~1613. Note, however, that this result applies to the Cepheid distance modulus derived from a multi--band fit. It does not necessarily tell us anything about the effect of metallicity on an individual PL relation if reddening and metallicity are covariant/degenerate in selected bandpasses.

\subsection{Dispersion of Independent Measurements}
\label{sec:dispersion}

\begin{figure}
\begin{center}$
 \begin{array}{c c} 
  \includegraphics[width=40mm, angle=-90]{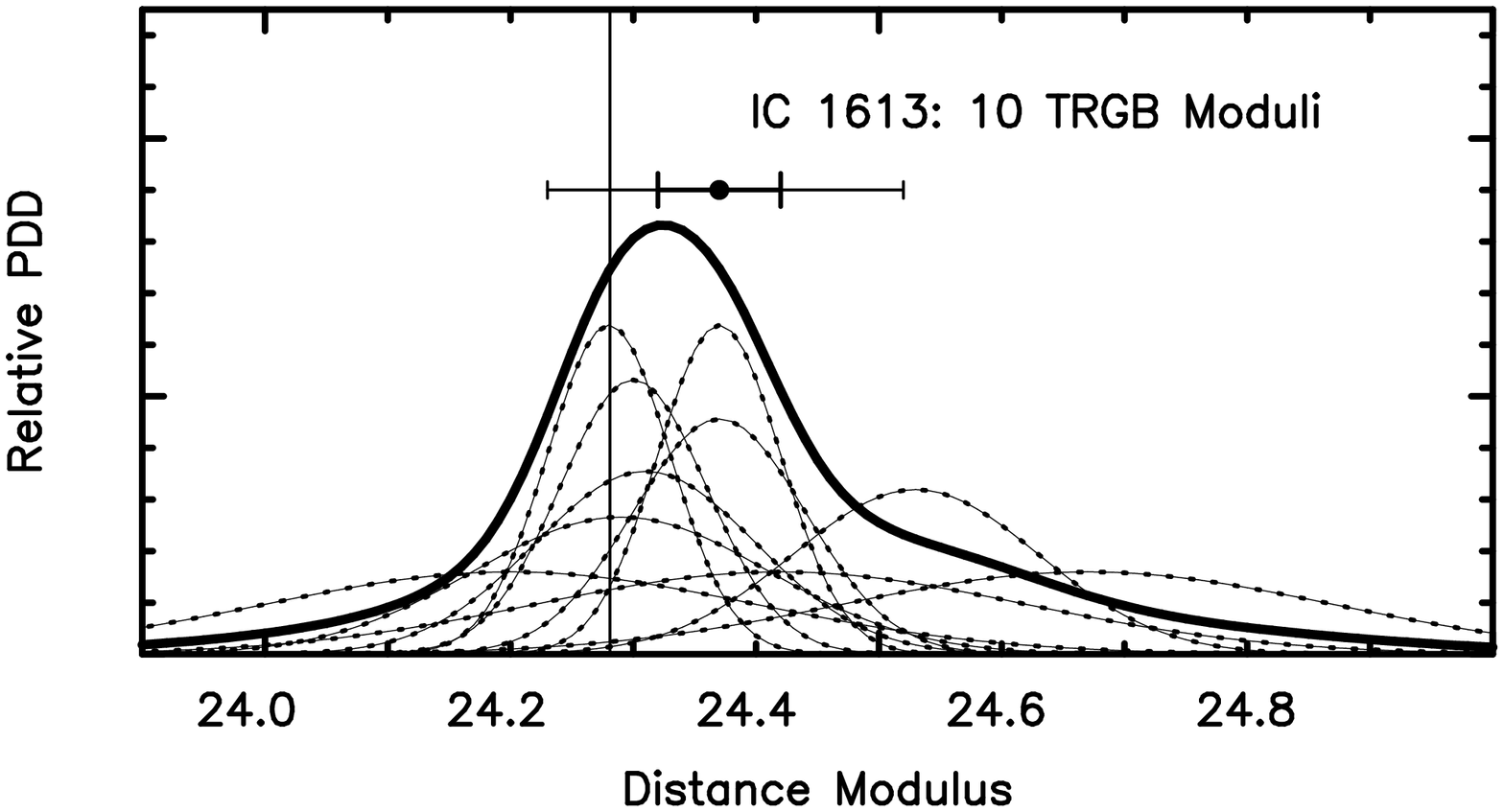} &
 \includegraphics[width=40mm, angle=-90]{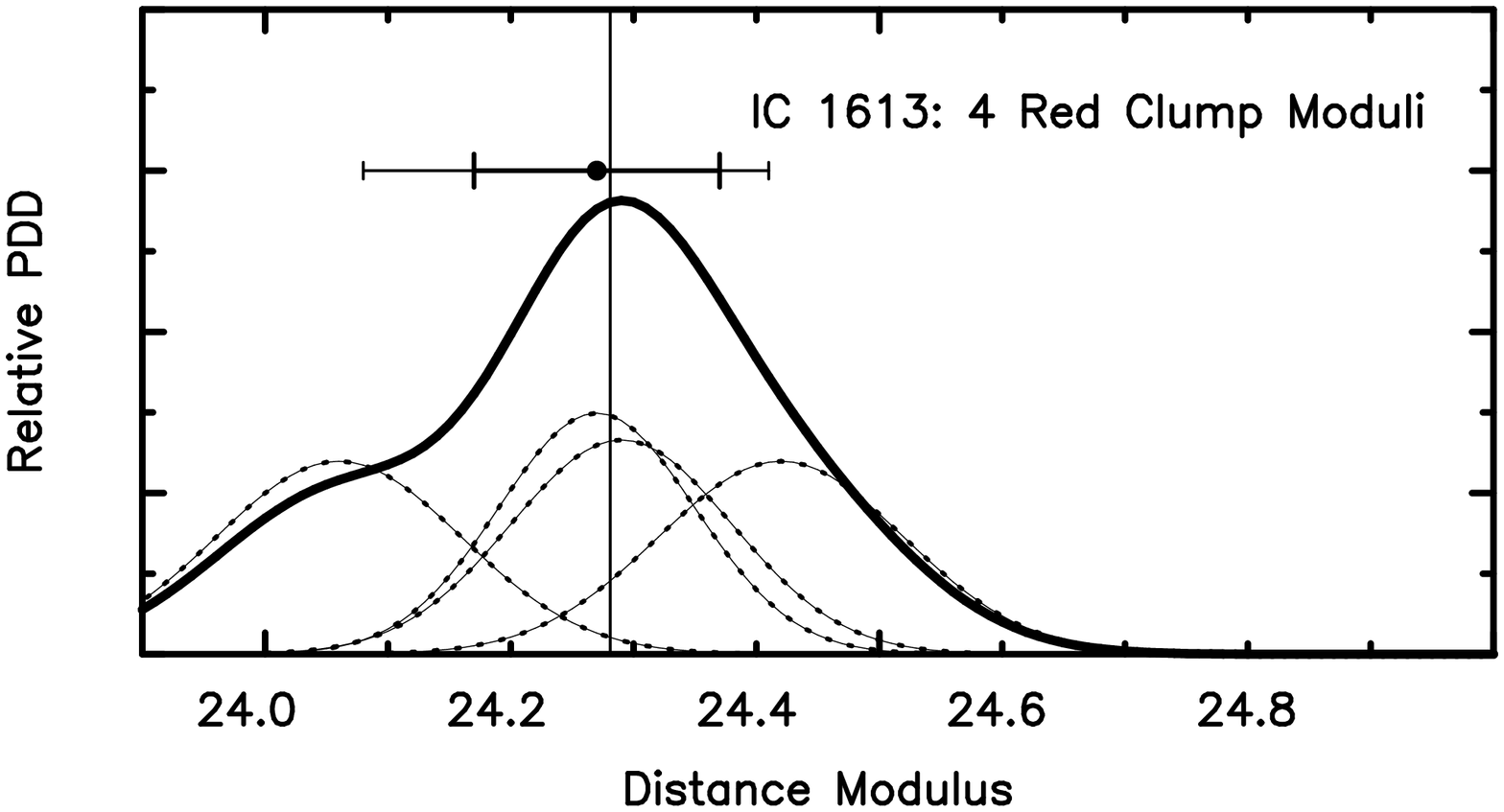} \\
 \mbox{\bf (a)} & \mbox{\bf (b)} \\
 \includegraphics[width=40mm, angle=-90]{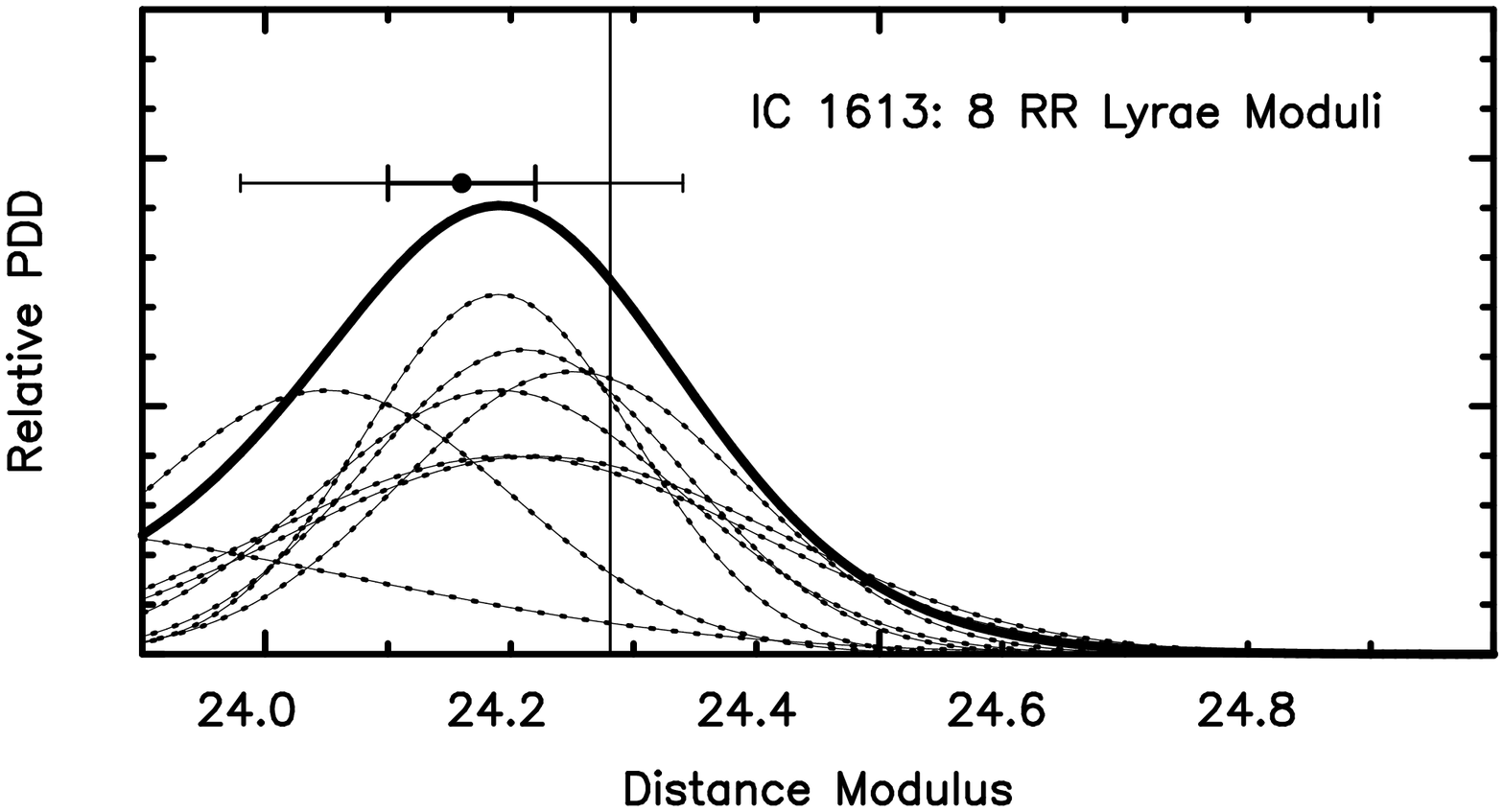} &
 \includegraphics[width=40mm, angle=-90]{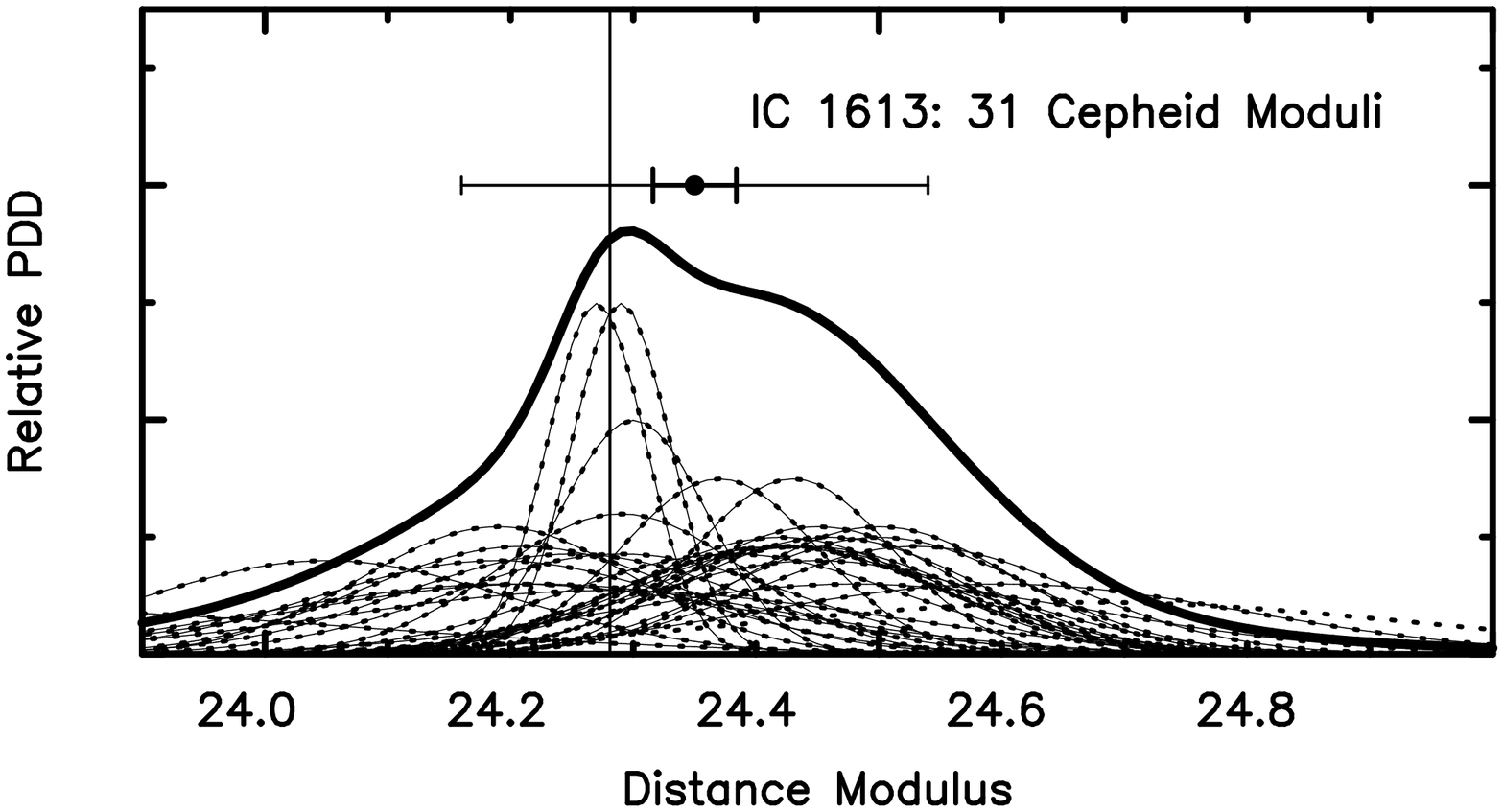}\\
 \mbox{\bf (c)} & \mbox{\bf (d)} \\
\end{array}$
\caption{A montage of individual comparisons of distance moduli to IC 1613 as published over the years and broken down into four major methods: the TRGB (a), the Red Clump (b), the RR Lyraes (c) and the Classical Cepheids (d). Individual distance determinations are shown as unit-area gaussians. The cumulative distribution is shown as the thick solid line. The median value is shown as a solid point with error bars. The larger error bars capture 68\% of the density around the median. The smaller error bar is the error on the mean. For consistent comparison, the vertical solid line marks the Cepheid distance determined in this paper.}
\label{fig:distance_comparison}
\end{center}
\end{figure}

In Figure~\ref{fig:distance_comparison} we make a graphical comparison of our newly determined Cepheid distance to IC~1613 with the published record of prior distance determinations as found in the December 2012 release of the compilation of redshift-independent distances in NED-D. No attempt has been made to put any of these distances onto a common zero point; the data therefore reflect a variety of adopted reddenings, zero points and wavelengths. We have however, subdivided the data down to a comparison of three major methods: the tip of the red giant branch (TRGB) method, the RR Lyraes, red clump stars and previous determinations also using Cepheids. The distance modulus determined above is shown as a solid vertical line in each of the plots. Individual determinations are shown as unit-area gaussians whose width is the published error on the mean. The solid line is the renormalized (Frequentist) sum of those individual probability density distributions (PDD); its peak is the mode of the published distribution. The filled circle (with horizontal error bars) is the median value of the PDD. The larger error bars capture 68\% of the density around the median. The smaller error bar is the error on the mean. 

On average, the (Population II) RR Lyrae variables are seen to give a slightly lower distance moduli than our (Population I) Cepheids. On the other hand, the (Population II) TRGB Method appears to give, on average, slightly larger moduli than our Cepheid modulus, although specific studies can be selected that agree exactly. The red clump moduli are few in number and widely scattered, although they do broadly agree with our Cepheid distance.  The largest number of distance determinations come from previously published studies of the Cepheids themselves. Here we compare 31 previous determinations and remark that the mode of this distribution is in good agreement with the latest value, although the range of values accumulated over the years is considerable.

In their Figure 16, B10 plot the distance moduli in the literature when corrected to a common LMC distance and common $E(B-V)$. The dispersion is somewhat reduced, but we note that their adopted $\mu_{LMC}$ is 18.515 mag (compare to the CHP value of 18.48 mag), and their adopted $E(B-V) = 0.025$ mag is lower than the $E(B-V) = 0.05$ mag we derive from our multi--band fit. It is clear from this, and Figure~\ref{fig:distance_comparison} that adopted reddening and the LMC distance are the dominant systematics in the determination of the distance to IC~1613. 

Fortunately, reducing these two systematics is entirely the domain of the CHP. Our distance ladder is tied to parallax measurements of MW Cepheids, and we virtually eliminate reddening in the mid--infrared. Our result, $\mu_{0} = 24.29 \pm 0.03_{stat} \pm 0.03_{sys}$ and its quoted errors reflect the reduced systematic uncertainty and increased precision in these values, showing the power of moving to the mid--infrared for Cepheid distance studies.

\subsection{Metallicity Effects in the Mid--Infrared}
\label{sec:metallicity}
IC~1613 has a metallicity of $[Fe/H]\approx -1$ (D01), significantly lower than the MW and LMC which were the two calibration galaxies for the CHP PL relations.  This makes it an ideal test--bed for searching for metallicity effects in the Cepheid PL. Initial tests for metallicity sensitivity in the mid--infrared were presented in \citet{2012ApJ...758...24F}, where we plotted the residual from the PL relation against spectroscopic metallicity for individual Cepheids. We found no significant effect at $-0.6 \leq [Fe/H] \leq 0.2$. 
Including IC~1613 in our studies increases the metallicity range of CHP Cepheids by a factor of two, so if a significant metallicity effect were present we should be able to detect it somewhere in this range. 

To test for a metallicity effect in IC~1613 we must use a different tactic from the one we applied in the MW and LMC. As we do not have metallicity measurements of the individual stars we must treat them as an ensemble. We assume that there is no effect on the PL slope and that any difference would manifest itself in the zero--point. Therefore, if composition does have an effect, we should find a different distance modulus than with an independent measure such as the TRGB or red clump. 

Correcting our $[3.6]$ and $[4.5]$ distances for extinction using $E(B-V) = 0.05 \pm 0.01$~mag we derive $\mu_{0,[3.6]} = 24.30 \pm 0.09$~mag and $\mu_{0,[4.5]} = 24.25 \pm 0.08$~mag, respectively. Both of these values are in excellent agreement with the TRGB distance from D01. The 4.5 $\mu$m distance modulus is slightly (0.05 mag) brighter than $\mu_{0,[3.6]}$ but still agrees to within one $\sigma$. As has been discussed previously \citep{2011AJ....142..192F, 2011ApJ...743...76S, 2012ApJ...759..146M}, we believe that the [4.5] band is unsuitable for distance measurements as it is contaminated by the temperature and metallicity sensitive CO band-head at 4.6 $\mu$m. Henceforth, all references to our mid--infrared Cepheid distance pertain solely to the 3.6 $\mu$m measurement.

The consistency of our [3.6] distance modulus with the TRGB, red clump and multi--band distances shows again that metallicity is not significantly impacting the distance measurements over the range $-1.0 \leq [Fe/H] \leq 0.2$. Therefore we conclude that there is no effect of metallicity on the 3.6 $\mu$m Cepheid PL relation zero--point at the level of $\pm 0.09$~mag. The mid--infrared measurement of the Hubble constant needs no adjustment for metallicity effects.

\section{Summary and Conclusions}
\label{sec:conclusions}
We performed a multi--epoch survey of IC~1613 using \textit{Spitzer} in the mid--infrared and the new FourStar camera on Magellan in the near--infrared. The photometric catalogs were matched to the OGLE Cepheid catalog to locate the Cepheids. Mean--light magnitudes were obtained for each star and PL relations were constructed in the $J$, $H$, $K_{s}$, $[3.6]$ and $[4.5]$ bands, from which distance moduli were derived. Using the 3.6 $\mu$m PL relation, where the effects of reddening are minimized, we measure the true distance modulus of IC~1613 as $\mu_{0,3.6} = 24.30 \pm 0.09_{stat} \pm 0.03_{sys}$ mag. This is entirely consistent with the independent TRGB  and red clump distance moduli derived in \citet{2001ApJ...550..554D}.

In addition to the single--band mid--infrared distance we have used near--infrared data from FourStar and archival optical data (corrected to an LMC distance of $\mu_{LMC} = 18.48 \pm 0.03$~mag) to derive a nine--band fit to measure the reddening and distance modulus of IC~1613. We find $E(B-V) = 0.05 \pm 0.01$ and $\mu_{0} = 24.29 \pm 0.03_{stat} \pm 0.03_{sys}$~mag. 

Finally, we have shown that as the mid--infrared Cepheid distance agrees with the TRGB distance, there must be no significant metallicity effect on the PL relation in the range $-1.0 \leq [Fe/H] \leq 0.2$. This removes any uncertainty in the CHP distance scale due to metallicity effects in the Cepheid calibration, significantly reducing the uncertainty in the CHP mid--infrared determination of the Hubble constant.

\acknowledgments
We thank the referee for their thoughtful response, which improved the content and style of the paper. 

This research made use of the NASA/IPAC Extragalactic database (NED) which is operated by the Jet Propulsion Laboratory, 
California Institute of Technology, under contract with the National Aeronautics and Space Administration.

This research made use of APLpy, an open-source plotting package for Python hosted at http://aplpy.github.com


%% in parentheses, after the keyword, but they will not be verified.

{\it Facilities:} \facility{Spitzer}, \facility{Magellan:Baade}



\appendix

\section{Fitting Template Light Curves}
\label{sec:app_template_fitting}
The amplitudes of Cepheid light curves in the near--infrared are much smaller than at optical wavelengths \citep[approximately $1/3$ to $2/3$ of the $V$ or $I$ band amplitudes --- see Table 2 of][]{2005PASP..117..823S}. However, they still reach levels around 0.5~mag which, combined with non--uniform sampling, can significantly affect the mean--light magnitude derived from a straight average. The effect of small numbers of non--uniform observations can be negated using template fitting, as was first shown by \citet{1988ApJ...326..691F} and later elaborated upon by \citet{2005PASP..117..823S}. In that paper they made template light curves in the $J$, $H$, and $K_{s}$ bands that could be scaled and phased using a complete $V$ or $I$ band light curve and a single near--infrared observation.

The technique was used by the Araucaria project to derive higher accuracy mean--light magnitudes from single observations in the $J$ and $K$ bands. They successfully applied the method to Cepheids in IC~1613 \citep{2006ApJ...642..216P}, obtaining PL relations in each band from either one or two observations per star. Here we test this method to derive mean--light magnitudes for our near--infrared Cepheids.

The most fundamental piece of information in the template fitting technique is the period of the Cepheid. From this and the $V$ or $I$ light curve the time of maximum light in the reference (optical) band is predicted. The phase--lag between the reference band and the near--IR band is known and can be used to predict the time of maximum light in each of the near--IR bands. The amplitude of the light curve is scaled to the amplitude of the reference light curve. The template is then fit to each near--IR observation individually and the mean--light magnitude is calculated; a weighted mean of these values gives the best estimate of the mean--light magnitude of the Cepheid. 

It is imperative to have highly precise periods for the Cepheids; if the time of maximum light is computed incorrectly then the relative phases of each data point will be erroneous and the mean--light magnitude will be incorrect. As the observations in \citet{2006ApJ...642..216P} were taken several years after the data for the reference light curves they took three more $V$ band observations contemporaneously with the near--IR data. This allowed them to refine the periods and define the time of maximum light ($\phi = 0$) more accurately. The periods typically changed by 0.1 to 0.5\%, but over 10 years this $\Delta P$ is sufficient to shift the time of maximum light by as much as $\phi = \pm 0.5$ compared to the original estimate. 

\begin{figure}
\begin{center}$ 
 \begin{array}{c} 
 \includegraphics[width=110mm]{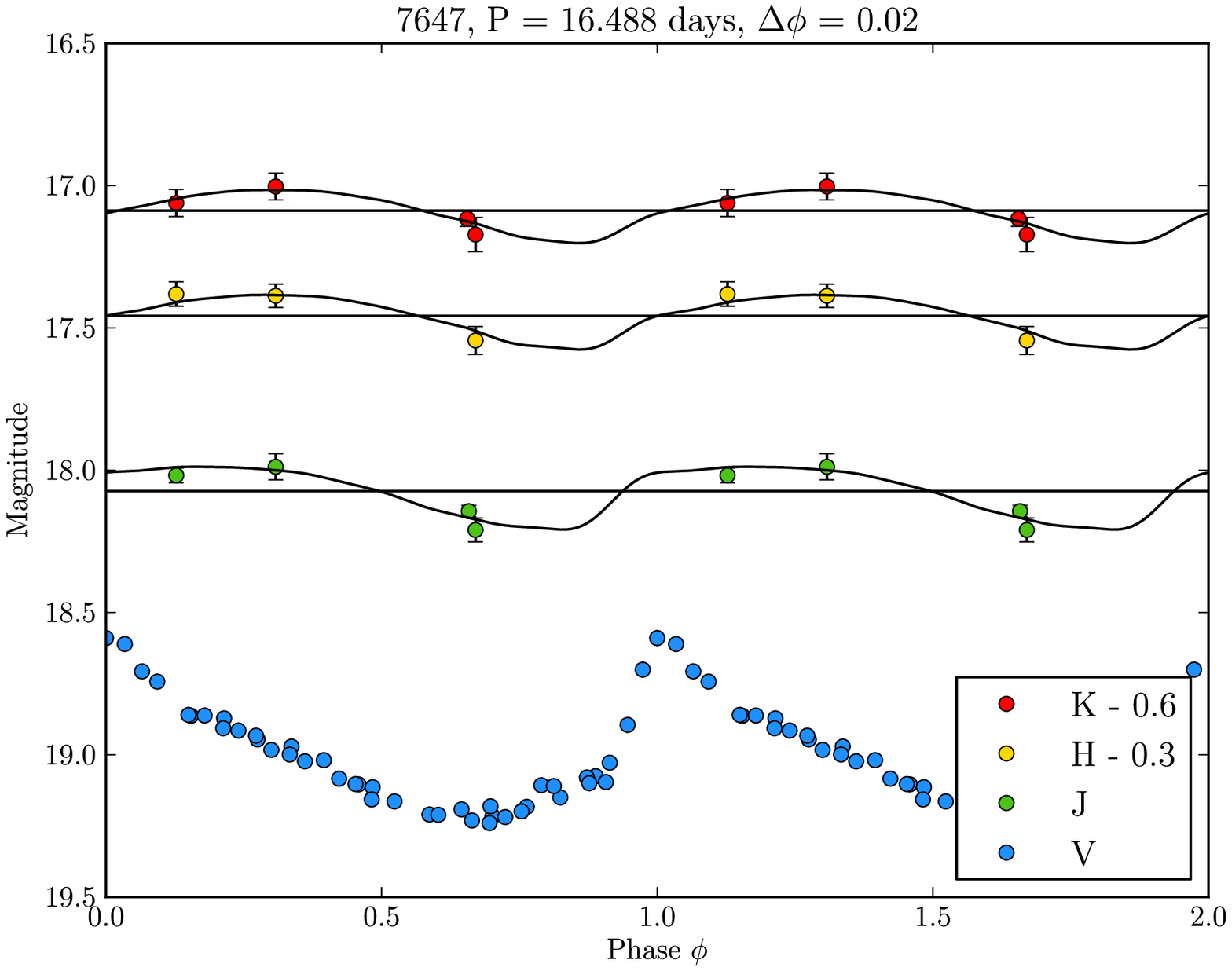} \\ [1cm]
  \includegraphics[width=110mm]{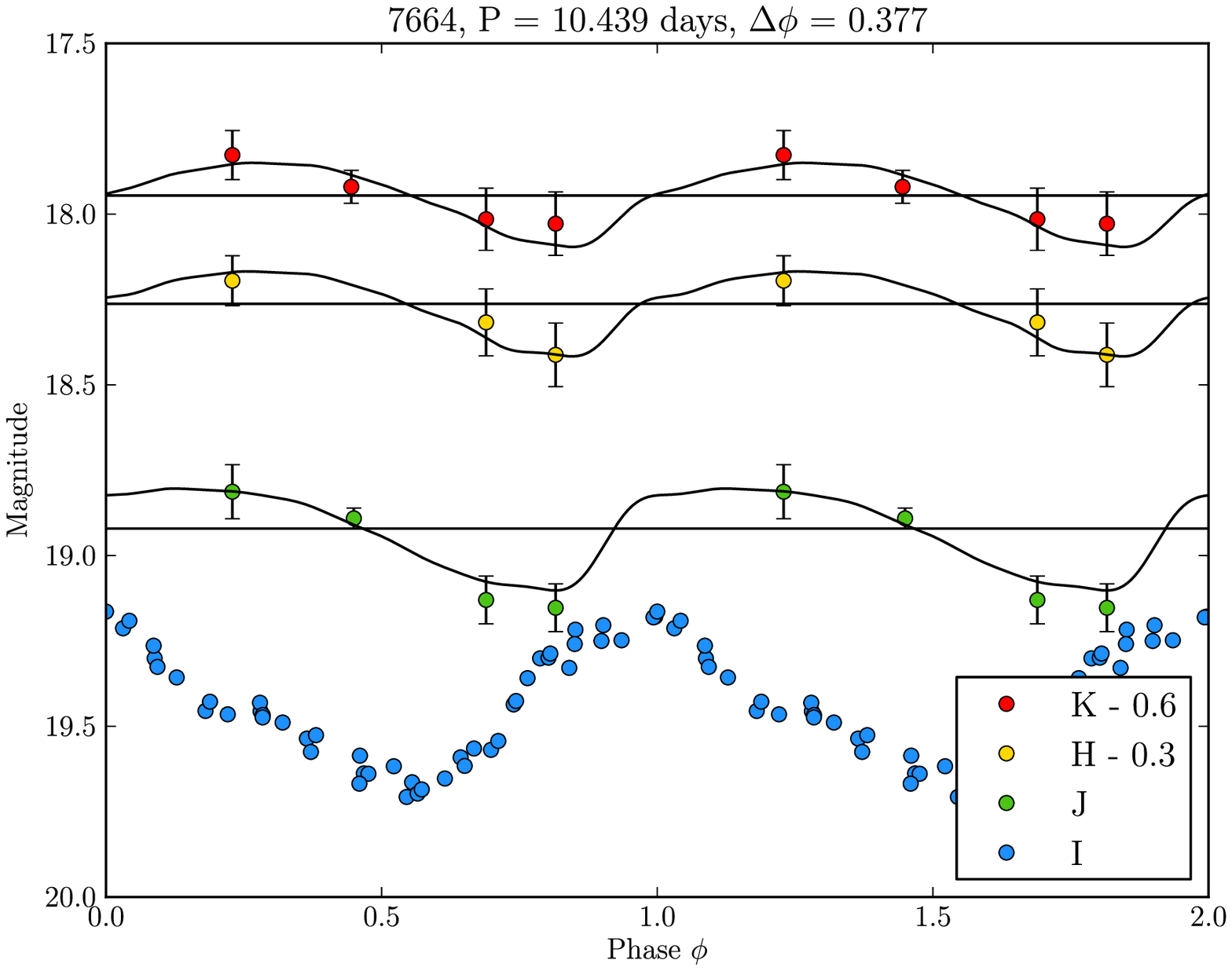} \\
\end{array}$
\caption{Light curves fit using the template method from \citet{2005PASP..117..823S}. A phase shift, $\Delta \phi$ was incorporated into the algorithm to account for possible period changes, and hence deviations from the predicted time of maximum light in the near--IR bands. The $J$ and $K_{s}$ light curves also contain points from \citet{2006ApJ...642..216P} where available. The Cepheid IDs use the OGLE numbering scheme.}
\label{fig:example_lcs}
\end{center}
\end{figure}

We do not have contemporaneous optical data to further refine the periods. We adopted the periods from \citet{2006ApJ...642..216P} where they were available and \citet{2001AcA....51..221U} in all other cases. To account for the less accurate periods the fitting algorithm was altered to allow for a phase shift. The best--fit phase shift was calculated by stepping through the possible shifts with a step size of $\delta \phi = 0.001$ and minimizing the residuals of the points around the template light curve. Example light curves for two Cepheids are shown in Figure~\ref{fig:example_lcs}. The phase shifts were found to be anywhere in the range $-0.4 \leq \Delta \phi \leq 0.4$.  To reiterate, shifts of this size could be induced by a period change of less than 1\% over 10 years. At this point it is clear that we do not know the periods of the Cepheids well enough to predict the time of maximum light to the required degree of accuracy, and so cannot determine the phase of any of our data points with a high degree of confidence. 

To confirm these thoughts the PL relations were plotted using the template mean--light magnitudes. The resulting apparent moduli showed marginal changes --- at the level of $1 \sigma$ --- but no significant differences. 

We conclude that although our knowledge of the periods of the Cepheids is good enough to derive a PL relation, it is not sufficient for determining the time of maximum light to the accuracy required for the template fitting technique. For the rest of this work we adopt the regular mean values for $J$, $H$, and $K_{s}$ as listed in Table~\ref{tab:cep_phot}.

%Appendix goes here



\label{references}
%\begin{thebibliography}{}
\bibliography{scowcroft_ic1613_2013}
%\bibitem[\protect\citeauthoryear{

%\bibitem[\protect\citeauthoryear{

%\end{thebibliography}

\clearpage

\begin{table}
\begin{center}
\begin{tabular}{ l  c c c c c } \\ \hline \hline
Date & Band & Exposure time & N$_{dither}$ & N$_{coadd}$ & Total exposure (s) \\  \hline
2011-09-09 & $J$ & 20.38 & 9 & 2 & 367 \\ 
 MJD = 55811 & $H$ & 8.733 & 9 & 4 & 314 \\
 & $K_{s}$ & 14.56 & 9 & 2 & 262 \\ \hline
 2011-10-04  & $J$ & 20.38 & 9 & 2 & 376 \\
 MJD = 55838& $H$ & 8.733 & 9 & 4 & 314 \\
 & $K_{s}$ & 14.56 & 9 & 2 & 262 \\ \hline
2011-11-03 & $J$  & 20.38 & 9 & 2 & 367 \\
 MJD = 55868 & $H$ & 5.822 & 9 & 6 & 314 \\
& $K_{s}$ & 5.822 & 9 & 6 & 314 \\ \hline
\end{tabular}
\caption{FourStar observations of IC~1613.}
\label{tab:fourstar_obs}
\end{center}
\end{table}

\begin{table}
\begin{center}
\begin{tabular}{ l c c c } \\ \hline \hline
Date & Average HMJD  & Block & Epoch\\ \hline
2010-01-26 & 55222.16 & 1 & 1 \\
2010-02-06 & 55234.01 & 1 & 2 \\
2010-02-14 & 55241.04 & 1 & 3 \\
2010-02-25 & 55252.41 & 1 & 4 \\ \hline
2010-08-20 & 55428.68 & 2 & 5 \\
2010-08-28 & 55436.73 & 2 & 6 \\
2010-09-07 & 55446.82 & 2 & 7 \\
2010-09-17 & 55456.03 & 2 & 8 \\ \hline
2011-02-03 & 55595.95 & 3 & 9 \\
2011-02-13 & 55605.14 & 3 & 10 \\
2011-02-23 & 55615.08 & 3 & 11 \\
2011-03-05 & 55625.57 & 3 & 12 \\
\hline
\end{tabular}
\caption{IRAC observations of IC~1613.}
\label{tab:observations}
\end{center}
\end{table}

\begin{landscape}
\begin{deluxetable}{l c c c c c c c c c c c c c c }
\tabletypesize{\scriptsize}
\tablecolumns{16}
\tablewidth{0pc}
\tablecaption{Near-- and mid--infrared mean magnitudes of Cepheids found in IC~1613. \label{tab:cep_phot}
}
\tablehead{
\colhead{OGLE ID} & \colhead{Period} & \colhead{RA} & \colhead{Dec} & \colhead{$J$} & \colhead{$\sigma_{J}$} &\colhead{$H$} & \colhead{$\sigma_{H}$} &\colhead{$K_{s}$} & \colhead{$\sigma_{K_{s}}$} &\colhead{[3.6]} & \colhead{$\sigma_{[3.6]}$} & \colhead{[4.5]} & \colhead{$\sigma_{[4.5]}$}  & S71 \\
  & \colhead{(days)} & \colhead{(hh:mm:ss)} & \colhead{(dd:mm:ss)} &\colhead {(mag)} & \colhead {(mag)} & \colhead {(mag)} & \colhead {(mag)} & \colhead {(mag)} &\colhead {(mag)} &\colhead {(mag)} &\colhead {(mag)} &\colhead {(mag)} &\colhead {(mag)} & }
\startdata
V22\tablenotemark{a, b} & 123.880 & 1:05:00.701 & +02:10:48.60 & 15.878 & 0.004 & 15.644 & 0.003 & 15.454 & 0.002 & 15.369 & 0.009 & 15.457 & 0.017 & V22 \\
11446\tablenotemark{a} & 41.630 & 1:04:59.740 & +02:05:28.30 & 17.139 & 0.012 & 16.970 & 0.009 & 16.866 & 0.007 & 16.600 & 0.014 & 16.610 & 0.018 & V20  \\
736 & 23.45 & 1:04:32.130 & +02:05:01.90 & ... & ... & ... & ... & ... & ... & 17.903 & 0.162 & 17.325 & 0.043 & V2 \\
7647\tablenotemark{a} & 16.540 & 1:04:37.700 & +02:09:08.40 & 18.052 & 0.020 & 17.916 & 0.016 & 17.854 & 0.014 & 17.656 & 0.032 & 17.694 & 0.031 & ... \\
13738 & 16.37 & 1:05:02.810 & +02:10:35.10 & 18.440 & 0.027 & 18.210 & 0.019 & 18.025 & 0.014 & 18.040 & 0.021 & 18.199 & 0.021 & V18  \\
7664\tablenotemark{a} & 10.450 & 1:04:41.420 & +02:08:24.20 & 19.038 & 0.042 & 18.825 & 0.032 & 18.726 & 0.027 & 18.585 & 0.025 & 18.637 & 0.039 & V16 \\
926\tablenotemark{a} & 9.402 & 1:04:33.590 & +02:07:45.60 & 19.016 & 0.041 & 18.838 & 0.029 & 18.784 & 0.025 & 18.597 & 0.029 & 18.541 & 0.045 & V06 \\
11589 & 8.409 & 1:04:51.510 & +02:05:33.50 & 19.490 & 0.075 & 19.291 & 0.050 & 19.245 & 0.043 & 18.404 & 0.056 & 18.504 & 0.0585 & V34  \\
13808 & 7.557 & 1:04:59.740 & +02:08:43.10 & 19.617 & 0.069 & 19.332 & 0.044 & 19.262 & 0.038 & ... & ... & ... & ... &  ... \\
13759 & 7.333 & 1:04:52.510 & +02:08:04.80 & 19.491 & 0.068 & 19.331 & 0.048 & 19.298 & 0.043 & 18.336 & 0.068 & 18.451 & 0.071 & V7  \\
18905 & 6.766 & 1:05:06.310 & +02:12:33.90 & 19.734 & 0.050 & 19.481 & 0.038 & 19.441 & 0.033  & 18.989 & 0.079 & 19.310 & 0.104 &  ... \\
13943\tablenotemark{a} & 6.751 & 1:04:51.670 & +02:10:55.00 & 19.407 & 0.091 & 19.217 & 0.064 & 19.134 & 0.054 & 18.990 & 0.041 & 19.101 & 0.046 & V24 \\
3732 & 6.669 & 1:04:40.210 & +02:01:24.80 & ... & ... & ... & ... & ... & ... & 19.250 & 0.041 & 19.117 & 0.080 & V27 \\
5037 & 6.31 & 1:04:49.140 & +02:07:20.20  & 20.149 & 0.109 & 19.850 & 0.080 & 19.790 & 0.072 & 19.140 & 0.079 & 18.958 & 0.103 & ... \\
3722 & 5.818 & 1:04:43.830 & +02:01:04.70 & ... & ... & ... & ... & ... & ... & 19.734 & 0.074 & 19.317 & 0.082 & V26 \\
13911 & 5.717 & 1:04:51.600 & +02:10:10.50 & 19.911 & 0.081 & 19.688 & 0.055 & 19.638 & 0.050 & 19.481 & 0.064 & 19.401 & 0.085 & V17 \\
13780 & 5.58 & 1:04:56.250 & +02:08:21.60 & 19.965 & 0.087 & 19.717 & 0.060 & 19.684 & 0.053 & 19.501 & 0.058 & 19.407 & 0.082 & V9 \\
4875 & 5.138 & 1:04:48.980 & +02:05:37.10 & 19.720 & 0.079 & 19.579 & 0.060 & 19.542 & 0.051 & 19.330 & 0.041 & 19.283 & 0.058 & V14 \\
15696 & 5.012 & 1:04:50.930 & +02:14:30.60 & ... & ... & ... & ... & ... & ... & 19.426 & 0.072 & 19.645 & 0.078 & ... \\
15670 & 4.849 & 1:04:53.290 & +02:13:30.60 & ... & ... & ... & ... & ... & ... & 19.456 & 0.061 & 19.312 & 0.101 & V13 \\
14287 & 4.365 & 1:05:01.050 & +02:09:11.80 & 20.484 & 0.104 & 20.252 & 0.079 & 20.202 & 0.070 & 19.582 & 0.091 & 19.692 & 0.134 &   ... \\
13784 & 4.045 & 1:04:59.848 & +01:53:10.16 & ... & ... & ... & ... & ... & ... & 19.216 & 0.047 & 19.385 & 0.075 & V10 \\
6084 & 3.872 & 1:04:46.550 & +02:07:28.10 & ... & ... & ... & ... & ... & ... & 19.667 & 0.087 & 19.329 & 0.095 & ...  \\
\enddata
%\tablecomments{Cepheids from the OGLE and Sandage catalogs were located by position.}
\tablenotetext{a} {Detected by Freedman et al. (2009)}
\tablenotetext{b}{No OGLE ID}
\end{deluxetable}
\end{landscape}

%\begin{landscape}
\begin{deluxetable}{l c c c c c c}
%\tabletypesize{\scriptsize}
\tablecolumns{7}
\tablewidth{0pc}
\tablecaption{$J$ band time series photometry of Cepheids in IC~1613. \label{tab:j_phot}
}
\tablehead{
\colhead{OGLE ID} & \colhead{$J_{1}$\tablenotemark{a}} & \colhead{$\sigma_{J_{1}}$} & \colhead{$J_{2}$\tablenotemark{b}} & \colhead{$\sigma_{J_{2}}$} & \colhead{$J_{3}$\tablenotemark{c}} & \colhead{$\sigma_{J_{3}}$}\\
  & \colhead {(mag)} & \colhead {(mag)} & \colhead {(mag)} & \colhead {(mag)} & \colhead {(mag)} &\colhead {(mag)}}
\startdata
V22 & 15.709 & 0.008 & 15.754 & 0.007 & 16.192 & 0.007\\
11446 & 17.079 & 0.014 & 17.205 & 0.057 & 17.397 & 0.028\\
736 & ... & ... & ... & ... & ... & ...\\
7467 & 18.21 & 0.042 & 17.988 & 0.046 & 18.018 & 0.026\\
13738 & 18.67 & 0.079 & 18.454 & 0.038 & 18.36 & 0.044\\
7664 & 18.813 & 0.079 & 19.153 & 0.07 & 19.13 & 0.07\\
926 & 19.011 & 0.057 & 19.02 & 0.083 & 19.023 & 0.085\\
11589 & 19.354 & 0.14 & 19.457 & 0.126 & 19.654 & 0.126\\
13808 & 19.661 & 0.127 & 19.581 & 0.128 & 19.613 & 0.109\\
13759 & 19.497 & 0.13 & 19.459 & 0.113 & 19.519 & 0.111\\
18905 & 19.742 & 0.084 & 19.789 & 0.111 & 19.702 & 0.076\\
13943 & ... & ... & 19.407 & 0.091 & ... & ... \\
3732  & ... & ... & ... & ... & ... & ...\\
5037 & 20.056 & 0.18 & 19.975 & 0.16 & 21.408 & 0.268\\
3722 & ... & ... & ... & ... & ... & ...\\
13911 & 20.017 & 0.148 & 19.807 & 0.137 & 19.935 & 0.137\\
13780 & 19.957 & 0.175 & 19.849 & 0.137 & 20.12 & 0.146\\
4875 & 19.718 & 0.134 & 19.743 & 0.144 & 19.703 & 0.135\\
15696 & ... & ... & ... & ... & ... & ...\\
15670 & ... & ... & ... & ... & ... & ...\\
14287 & ... & ... & ... & ... & ... & ...\\
13784 & ... & ... & ... & ... & ... & ...\\
6084 & ... & ... & ... & ... & ... & ...\\
\enddata
\tablenotetext{a} {MJD$_{1}$ - 2400000 = 55811}
\tablenotetext{b}{MJD$_{2}$ - 2400000= 55838}
\tablenotetext{c}{MJD$_{3}$ - 2400000= 55868}
\end{deluxetable}
%\end{landscape}

%\begin{landscape}
\begin{deluxetable}{l c c c c c c}
%\tabletypesize{\scriptsize}
\tablecolumns{7}
\tablewidth{0pc}
\tablecaption{$H$ band time series photometry of Cepheids in IC~1613. \label{tab:h_phot}
}
\tablehead{
\colhead{OGLE ID} & \colhead{$H_{1}$\tablenotemark{a}} & \colhead{$\sigma_{H_{1}}$} & \colhead{$H_{2}$\tablenotemark{b}} & \colhead{$\sigma_{H_{2}}$} & \colhead{$H_{3}$\tablenotemark{c}} & \colhead{$\sigma_{H_{3}}$}\\
  & \colhead {(mag)} & \colhead {(mag)} & \colhead {(mag)} & \colhead {(mag)} & \colhead {(mag)} &\colhead {(mag)}}
\startdata
V22 & 15.305 & 0.007 & 15.357 & 0.009 & 15.652 & 0.006\\
11446 & 16.622 & 0.034 & 16.82 & 0.019 & 16.803 & 0.027\\
736 & ... & ... & ... & ... & ... & ...\\
7467 & 17.844 & 0.049 & 17.687 & 0.041 & 17.681 & 0.043\\
13738 & 18.257 & 0.043 & 17.975 & 0.046 & 17.839 & 0.049\\
7664 & 18.495 & 0.073 & 18.712 & 0.093 & 18.617 & 0.098\\
926 & 18.67 & 0.077 & 18.737 & 0.071 & 18.641 & 0.069\\
11589 & 19.085 & 0.104 & 19.179 & 0.115 & 19.252 & 0.13\\
13808 & 19.232 & 0.101 & 19.162 & 0.102 & 19.127 & 0.097\\
13759 & 19.178 & 0.112 & 19.227 & 0.137 & 19.164 & 0.115\\
18905 & 19.328 & 0.133 & 19.177 & 0.091 & 19.227 & 0.087\\
13943 & ... & ... & 19.058 & 0.09 & ... & ...\\
3732 & ... & ... & ... & ... & ... & ...\\
5037 & 19.637 & 0.18 & 19.554 & 0.152 & ... & ...\\
3722 & ... & ... & ... & ... & ... & ...\\
13911 & 19.702 & 0.154 & 19.474 & 0.123 & 19.474 & 0.123\\
13780 & 19.511 & 0.122 & 19.448 & 0.156 & 19.67 & 0.176\\
4875 & 19.449 & 0.174 & 19.361 & 0.175 & 19.416 & 0.142\\
15696 & ... & ... & ... & ... & ... & ...\\
15670 & ... & ... & ... & ... & ... & ...\\
14287 & 20.023 & 0.283 & 20.073 & 0.215 & 19.958 & 0.168\\
13784 & ... & ... & ... & ... & ... & ...\\
6084 & ... & ... & ... & ... & ... & ...\\
\enddata
\tablenotetext{a} {MJD$_{1}$ - 2400000 = 55811}
\tablenotetext{b}{MJD$_{2}$ - 2400000= 55838}
\tablenotetext{c}{MJD$_{3}$ - 2400000= 55868}
\end{deluxetable}
%\end{landscape}

%\begin{landscape}
\begin{deluxetable}{l c c c c c c}
%\tabletypesize{\scriptsize}
\tablecolumns{7}
\tablewidth{0pc}
\tablecaption{$K_{S}$ band time series photometry of Cepheids in IC~1613. \label{tab:k_phot}
}
\tablehead{
\colhead{OGLE ID} & \colhead{$K_{S1}$\tablenotemark{a}} & \colhead{$\sigma_{K_{S1}}$} & \colhead{$K_{S2}$\tablenotemark{b}} & \colhead{$\sigma_{K_{S2}}$} & \colhead{$K_{S3}$\tablenotemark{c}} & \colhead{$\sigma_{K_{S3}}$}\\
  & \colhead {(mag)} & \colhead {(mag)} & \colhead {(mag)} & \colhead {(mag)} & \colhead {(mag)} &\colhead {(mag)}}
\startdata
V22 & 15.159 & 0.005 & 15.189 & 0.006 & 15.532 & 0.007\\
11446 & 16.539 & 0.025 & 16.79 & 0.019 & 16.753 & 0.019\\
736 & ... & ... & ... & ... & ... & ...\\
7467 & 17.772 & 0.06 & 17.603 & 0.047 & 17.661 & 0.048\\
13738 & 18.161 & 0.048 & 17.841 & 0.041 & 17.741 & 0.028\\
7664 & 18.427 & 0.072 & 18.628 & 0.093 & 18.615 & 0.091\\
926 & 18.635 & 0.07 & 18.732 & 0.078 & 18.58 & 0.104\\
11589 & 19.019 & 0.148 & 19.119 & 0.148 & 19.236 & 0.138\\
13808 & 19.147 & 0.218 & 19.082 & 0.112 & 19.064 & 0.116\\
13759 & 19.16 & 0.182 & 19.207 & 0.167 & 19.149 & 0.162\\
18905 & 19.262 & 0.129 & 19.337 & 0.127 & 19.317 & 0.122\\
13943 & ... & ... & 18.963 & 0.098 & ... & ...\\
3732 & ... & ... & ... & ... & ... & ...\\
5037 & 19.556 & 0.168 & ... & ... & ... & ...\\
3722 & ... & ... & ... & ... & ... & ...\\
13911 & 19.46 & 0.194 & 19.437 & 0.208 & 19.426 & 0.203\\
13780 & 19.566 & 0.183 & 19.637 & 0.187 & 19.56 & 0.183\\
4875 & 19.459 & 0.171 & 19.497 & 0.179 & 19.405 & 0.161\\
15696 & ... & ... & ... & ... & ... & ...\\
15670 & ... & ... & ... & ... & ... & ...\\
14287 & ... & ... & 19.996 & 0.219 & 20.063 & 0.216\\
13784 & ... & ... & ... & ... & ... & ...\\
6084 & ... & ... & ... & ... & ... & ...\\
\enddata
\tablenotetext{a} {MJD$_{1}$ - 2400000 = 55811}
\tablenotetext{b}{MJD$_{2}$ - 2400000= 55838}
\tablenotetext{c}{MJD$_{3}$ - 2400000= 55868}
\end{deluxetable}
%\end{landscape}

\begin{table}
\begin{center}
\begin{tabular}{ l c c c} \\ \hline \hline
Band & Zero--point (mag)\tablenotemark{a}  & Standard deviation & $\mu$ (mag)\\ \hline
$J$ & $19.07 \pm 0.05$ & 0.206 & $24.35 \pm 0.05$\tablenotemark{b} \\
$H$ & $ 18.69\pm 0.04$ & 0.165 & $24.30 \pm 0.04$\tablenotemark{b} \\
$K_{s}$ & $ 18.64\pm 0.05 $ & 0.168 & $24.33 \pm 0.05$\tablenotemark{b} \\
$[3.6]$ & $18.51 \pm 0.08$ & 0.307 & $24.31 \pm 0.09$\tablenotemark{c} \\
$[4.5]$ & $18.50 \pm 0.07$ & 0.235 & $24.26 \pm 0.08$\tablenotemark{c} \\
\hline
\end{tabular}
\caption{Mid--infrared period--luminosity relation zero--points for unblended Cepheids in IC~1613. }
\tablenotetext{a}{PL relations took the form $M = a  (\log P - 1.0) + b$; the $a$ coefficients are taken from \citet{2004AJ....128.2239P} ($J$, $H$, $K_{S}$) and \citet{2011ApJ...743...76S} ($[3.6]$, $[4.5]$).}
\tablenotetext{b}{Distance moduli were calculated using the LMC PL relation zero--points and uncertainties from \citet{2004AJ....128.2239P}, and assuming $\mu_{0, LMC} = 18.48$. They have not been corrected for extinction.}
\tablenotetext{c}{Distance moduli were calculated using the MW PL relation zero--points and uncertainties from \citet{2012ApJ...759..146M}. They have not been corrected for extinction.}
\label{tab:pl_relations}
\end{center}
\end{table}

\clearpage

\clearpage

\clearpage

\end{document}